\documentclass{emulateapj}

\defcitealias{dubinski92}{D92}
\defcitealias{js02}{JS02}
\defcitealias{warren-etal92}{W92}
\defcitealias{faltenbacher-etal02}{F02}
\defcitealias{ke04}{KE04}
\defcitealias{hrh00}{HRH00}
\defcitealias{hbb05}{HBB05}

\newcommand{\lcdm}{{$\mathrm{\Lambda}$CDM}}
\newcommand{\hkmskpc}{\ensuremath{h~\mathrm{km~s^{-1}~kpc^{-1}}}}
\newcommand{\rvir}{\ensuremath{r_{\mathrm{vir}}}}

\newcommand{\hkpc}{\ensuremath{h^{-1}~\mathrm{kpc}}}
\newcommand{\hmpc}{\ensuremath{h^{-1}~\mathrm{Mpc}}}

\newcommand{\hmsun}{\ensuremath{h^{-1}~\mathrm{M_{\Sun}}}}

\newcommand{\vhat}[1]{\ensuremath{\mathbf{\hat{#1}}}}

\shorttitle{Alignment of \lcdm\ Halos}
\shortauthors{Bailin and Steinmetz}

\begin{document}

\title{Internal and External Alignment of the Shapes and Angular
	Momenta of \lcdm\ Halos}\label{alignment chapter}
\author{Jeremy Bailin\altaffilmark{1,2,3,4} and
Matthias Steinmetz\altaffilmark{1,2,5,6}}
\altaffiltext{1}{Steward Observatory, University of Arizona, 933 N Cherry Ave,
 Tucson, AZ  85721 USA}
\altaffiltext{2}{Astrophysikalisches Institut Potsdam, An der Sternwarte 16,
 D-14482 Potsdam, Germany}
\altaffiltext{3}{Swinburne University of Technology, Mail H39, PO Box 218,
 Hawthorn, VIC 3121 Australia}
\altaffiltext{4}{jbailin@astro.swin.edu.au}
\altaffiltext{5}{msteinmetz@aip.de}
\altaffiltext{6}{David and Lucile Packard Fellow}

\begin{abstract}
We investigate how the shapes and angular momenta of galaxy and group
mass dark matter halos in a \lcdm\ $N$-body simulation are correlated
internally, and how they are aligned with respect to the
location and properties of surrounding halos.
We explore these relationships down to halos of much lower
mass ($10^{11}~\hmsun$) than previous studies.
The halos are triaxial, with $c/a$ ratios of $0.6\pm0.1$
and a mean two-dimensional projected ellipticity of $\left<e\right>=0.24$.
More massive halos are more flattened.
The axis ratios rise out to 0.6~\rvir, beyond which they drop.
The principal axes, in particular the minor axes,
are very well aligned within 0.6~\rvir.
High mass halos show particularly strong internal alignment.
The angular momentum vectors are also reasonably well aligned except
between the very outermost and very innermost regions of the halo.
The angular momentum vectors tend to align with the minor axes,
with a mean misalignment of $\sim 25\degr$,
and lie perpendicular to the major and intermediate axes.
The properties of a halo at 0.4~\rvir\ are quite characteristic
of the properties at most other radii within the halo.
There is a very strong tendency for the minor axes of halos to
lie perpendicular to large scale filaments,
and a much weaker tendency
for the major axes to lie along the filaments.
This alignment extends to much larger separations for group and
cluster mass halos than for galaxy mass halos.
As a consequence, the intrinsic alignments of galaxies are likely
weaker than previous predictions, which were based on the shapes of cluster
mass halos.
The angular momenta of the highest concentration halos tend to point toward
other halos.
The angular momenta of galaxy mass halos point parallel to
filaments, while those of group and cluster mass halos
show a very strong tendency to point perpendicular to the filaments.
This suggests that group and cluster mass halos acquire most of their angular
momentum from major mergers along filaments, while the accretion
history of mass and angular momentum onto galaxy mass halos has
been smoother.
\end{abstract}


\section{Introduction}

The three-dimensional structure of the dark matter halos that host galaxies,
groups, and clusters is an important aspect of their nature that can provide
insight into their formation and affect the luminous structures within.
The orientation of
the halo shapes and angular momenta, both internally
and with respect to surrounding halos, provide important constraints
on other studies of galaxy formation and evolution.

Halos formed in cosmological simulations are generally not spherical,
but have an ellipsoidal shape.
There have been several studies of the shapes of halos in
low resolution $N$-body simulations based on the
standard cold dark matter (CDM)
paradigm (\citealp{frenk-etal88,dc91}; \citealp{warren-etal92}
(hereafter \citetalias{warren-etal92}); \citealp{cl96}).
These studies have found that halos are usually triaxial, with a
preference for prolate figures at small radii and more oblate figures
at large radii, and have minor-to-major axis ratios ranging from 0.3
to almost unity. \citet{dc91}, in simulations of the formation
of isolated galaxies where the effects of the external tidal field
were prescriptively superimposed, found that the projected
two-dimensional ellipticities peak around $e=0.5$, where
\begin{equation}\label{ellipticity definition}
e \equiv 1 - q,
\end{equation}
for a projected axis ratio of $q$.
\citetalias{warren-etal92} found that while this holds in the
inner regions, the location of the peak falls to $e=0.25$ beyond
50~kpc. The behaviour of the axis ratios with radius is controversial;
\citet{dc91} and \citetalias{warren-etal92} find that the axis ratios increase
(become more spherical) with radius, while \citet{frenk-etal88} and
\citet{cl96} find that they decrease with radius.
\citetalias{warren-etal92} have also studied the internal
alignment of the ellipsoid principal axes. They
found that both the major and minor axes
of halos are extremely well aligned out to 40~kpc.

More recently, several authors have performed large high resolution
simulations 
using the currently-favoured \lcdm\ cosmology
(\citealp{bullock02}; \citealp{js02} (hereafter \citetalias{js02});
\citealp{ke04} (hereafter \citetalias{ke04}); \citealp{hbb05}
(hereafter \citetalias{hbb05})).
\citet{bullock02} finds that the $c/a$ axis ratios are a strong
function of halo mass,
and range from 0.55 at $10^{14}~\hmsun$ to 0.7 around
$10^{12}~\hmsun$, with a distribution that is peaked but
has a large tail to small axis ratios.
He also finds that the inner 30~\hkpc\ of halos are more spherical
than the outer regions, i.e.~the axis ratios decrease with radius.
On the other hand,
\citetalias{js02} find $c/a$ axis ratios
that increase with radius, decrease slightly with mass,
and are well fit by a Gaussian centred at $c/a=0.55$ with a
width of 0.11. They find that the major axes of halos are relatively
well aligned --- typically $\cos\theta_{11} \sim 0.8$, where
$\theta_{11}$ is the angle between the major axis at
small or large radius compared to that at an intermediate radius.
The alignment of the middle axes
is somewhat poorer, but \citetalias{js02} argue that this is due
to the inclusion of nearly prolate halos whose axes are degenerate and therefore
not well determined.
\citetalias{ke04} find more spherical shapes
in their very large sample of high mass halos
($M > 3 \times 10^{14}~\hmsun$),
possibly due to the spherical outer boundary they impose,
with $c/a \approx 0.65$.
\citetalias{hbb05} measure the intermediate-to-major axis ratio $b/a$
of cluster mass halos
and find a mean of 0.67, with a mass
dependence similar to that found by \citet{bullock02}, and
axis ratios that decrease with radius.

The shapes of dark matter halos can have important observational
consequences
\citep[for a good review, see][]{sackett99}.
On galactic scales, they can affect the coherence of tidal streams.
Some authors have claimed that the thinness of the tidal
stream associated with the Sagittarius dwarf spheroidal indicates
that the halo of of the Milky Way is nearly spherical, with $c/a \gtrsim 0.8$
\citep{ibata-etal01,jlm05,martinez-delgado-etal04}.
However, more recent studies suggest that the material that
makes up the stream was stripped from the satellite too recently
to have had time to undergo differential precession,
which severely weakens
the constraints on the halo shape \citep{helmi04a,helmi04b}.
\citet{helmi04b} even claims that the stream is best fit assuming a prolate
halo elongated perpendicular to the disk with $c/a \approx 0.6$.
\citet*{ljm05} also find that the velocities of stars in the leading
stream can only be fit with a prolate halo, but that the precession
of the leading stream
with respect to the trailing stream can only be fit with an oblate
halo.
These contradictory results suggest that evolution of the satellite
orbit or other effects of a live Milky Way potential (as opposed to
the static potential that has been used in all of these studies)
are important for determining the shape of the Milky Way halo
using the Sgr stream.
Other measures of the Milky Way ellipticity using the flaring of the
gas disk at large radius or the anisotropy of stellar velocities
suggest that the halo is oblate with a flattening of
$c/a \sim 0.8$ \citep{om00}.
The shapes of the
halos of external galaxies can be measured using the flaring of
the gas layer \citep{om00}, the projected shape of X-ray gas
\citep{buote-etal02}, or the kinematics of polar ring galaxies
\citep{sackett-etal94}.
These methods suggest that galaxy halos have a wide range of
flattenings from $c/a \sim 0.3$ -- $0.8$.
Assuming that the shape of the stellar halo traces
that of the dark matter halo,
the stacked images of edge-on disk galaxies in the Sloan
Digital Sky Survey are consistent with a mean $c/a = 0.6$
\citep{zwb04}.
A new method for measuring the shapes of external galaxy halos is
weak gravitational lensing.
By measuring the azimuthal
variation of the shear with respect to the position angle of
the visible lens galaxy \citep{nr00}, \citet{hyg04} detected
an average projected halo ellipticity of $\left< e \right> = 0.33$
for halos with an average mass of $8 \times 10^{11}~\hmsun$.
This detection also implies that the orientation of the visible and dark
mass in galaxies must be similar.
On group and cluster scales, X-ray observations and the
Sunyaev-Zeldovich effect
\citep{sz-effect}
can be directly used as a probe
of halo ellipticities \citep{ls03,ls04}.

The orientation of the angular momentum in halos has also been
studied in cosmological numerical simulations.
Early low-resolution studies \citep{be87,frenk-etal88,cl96}
gave conflicting results due to the difficulty of
measuring the direction of the angular momentum with few particles.
Other CDM studies (\citealp{dubinski92}; \citetalias{warren-etal92})
have found that the direction of the angular momentum at different
radii is usually the same, but that the distribution of alignments
has a tail that stretches all the way to anti-alignment.
This result is verified by recent high-resolution \lcdm\ simulations
\citep{bullock-etal01-angmom}.
CDM studies have also found that the angular momentum is most
often aligned with the minor axis and perpendicular to the major axis,
although there is some scatter (\citealp{dubinski92}; \citetalias{warren-etal92}).
This result has not yet been thoroughly tested in high-resolution
\lcdm\ simulations.

Internal misalignment of the angular momentum can have a number of
observational consequences. It may cause galactic warps
\citep{ob89,ds99,lopez-corredoira-etal02a,bailin-phd},
or manifest
itself in anisotropic distributions of the orbits of satellite galaxies
\citep{holmberg-effect,zaritsky-etal97,apc04,knebe-etal04}.

Going beyond individual halos, the shapes and angular momenta
of nearby halos can be correlated due to initial conditions
or dynamical evolution.
This subject has attracted increased interest recently due to
the emergence of weak gravitational lensing as a method to measure
the projected mass density in front of background galaxies.
Intrinsic correlations between the projected shapes of
luminous galaxies act as spurious background signals in weak lensing,
so predicting
their magnitude is important. At a more fundamental level, the degree
of correlation between structures can be tested against models,
and can inform our understanding of the origin of
halo shapes and angular momenta.

Measurements of halo alignments and correlations come from two
sources: cluster orientations and large galaxy surveys.
The study of the alignment of cluster orientations was pioneered
by \citet{binggeli82},
who used the locations of the constituent galaxies to determine
that the major axes of
clusters separated by less than 15~\hmpc\ tend to point toward each other.
While some authors have not found any such alignment
\citep[e.g.][]{sp85},
larger samples of both galaxies and clusters, along with improved
error estimates, have confirmed this result
\citep[e.g.][]{flin87,rk87,plionis94}.
Similar results are seen when using the major axis of the
brightest cluster galaxy as a measurement of the cluster
orientation \citep{lambas-etal90}.
Enhancements in galaxy counts along the major axis of
brightest cluster galaxies out to 15~\hmpc\ have also been detected
\citep{argyres-etal86,lgp88b,ml89}.
The cluster potential is better
probed by X-ray emitting gas \citep{sarazin86,ls03,ls04}.
While early studies using X-ray contours found no alignment
of clusters with the large scale structure \citep{umk89},
more and better data have confirmed that
the orientation of both the substructure and the main cluster potential
tends to point
toward neighbouring clusters \citep{wjf95,cmm02}.

While the principal axes of clusters can be determined from optical or
X-ray photometry, the angular momentum direction is very difficult
to determine.
In disk galaxies, on the other hand, the angular momentum direction of
the baryons,
presumed to be perpendicular to the orientation of the disk,
can be measured much more easily than the shape of
the dark matter halo.
Although there may be some misalignment between
the angular momentum of the baryons and dark matter 
\citep{vdb-etal02,ss04},
spiral galaxies still provide the best targets for detecting
angular momentum alignments and correlations.
Studies with small samples of spiral galaxies (less than a few hundred)
have generally found no correlation between the orientation
of the angular momentum and the large scale structure
(\citealp{cp90,han-etal95,cd99}; see however \citealp*{nas04}).
With a sample of 618 lenticular and disk galaxies in the local supercluster,
\citet{ko92} found that while the full sample was consistent
with an isotropic distribution of angular momenta, those galaxies
within 2~\hmpc\ of the supergalactic plane tend to have spin
vectors pointing in the plane, while those above or below the plane
tend to have spin vectors that point toward or away from the plane.
\citet{nas04} also find a clear excess of galaxies whose
angular momenta lie in the supergalactic plane.
Larger samples of galaxies provide further evidence of
alignments between spin and the large scale structure:
\citet*{pls00} have found that the spin directions of the 12,122 spiral
galaxies in the Tully catalog are positively correlated at separations
less than 3~\hmpc, while \citet{brown-etal02} have measured
intrinsic correlations between galaxy orientations
at a range of angular separations
in the $2\times 10^6$ galaxies of the SuperCOSMOS survey.

Linear tidal torque theory \citep{doroshkevich70,white84} can be
used to predict the directions of the angular momentum vectors
and their correlations with the surrounding structure
\citep{pls00,leepen00,leepen01,cnpt01,pdh02a,pdh02b}.
These studies have found that the angular momenta of halos
tend to lie perpendicular to the large scale structure, and that
the correlation of the halo spin vectors
with each other exists but is very weak.
While tidal torque theory performs reasonable well at predicting
the evolution of the magnitude of the angular momentum \citep{ssk00},
\citet{pdh02a} have tested the predictions of spin directions
against $N$-body simulations,
and found that the spin axes of $N$-body halos show significant misalignment
compared to the tidal torque predictions, with a mean misalignment
of $\sim 50\degr$ at $z=0$. Therefore, $N$-body simulations that
take the full non-linear dynamics into account are necessary.

Some early numerical work at predicting the intrinsic alignments
using simulations with power law or CDM power spectra found that
the major axes of cluster mass halos tend to point toward other
nearby clusters over scales of $\sim 15~\hmpc$, and that
there is a very weak tendency for the major axes to be 
correlated with each other over the same range of separations
\citep{be87,wvd91,vhvdw93,splinter-etal97}.
More recent high resolution $N$-body simulations in a
\lcdm\ cosmology have been studied to search for alignments
of the major axes (\citealp{ot00}; \citetalias{ke04}; \citetalias{hbb05}),
the angular momentum vectors
\citep{hn01}, or both (\citealp{faltenbacher-etal02};
hereafter \citetalias{faltenbacher-etal02}).
These studies have found a strong tendency for the major
axes of cluster mass halos to point toward other clusters
out to several tens (or even hundreds) of Mpc and
to correlate with each other out to 20~\hmpc.
The situation for the angular momentum is murkier.
\citet{hn01} have found that the angular momenta of halos tend to lie
parallel to the large scale structure,
while \citetalias{faltenbacher-etal02}
have found that they lie perpendicular to the large scale structure,
in agreement with the prediction from the linear theory \citep{leepen01}.
Correlations of halo angular momentum vectors with each other
appear weak at best.
Two groups have searched specifically for correlations
between the ellipticities of $z \approx 1$ galaxies in the
Virgo Consortium \lcdm\ model \citep{virgo-consortium}
in order to predict the effect on weak lensing signals.
\citet{cm00} assumed that the shapes of the galaxies were identical to
the shapes of the dark matter halos, while 
\citet{hrh00} (hereafter \citetalias{hrh00}) made
separate predictions about ellipticals,
which were assumed to share the shape of their dark matter halos, and
spirals, which were assumed to lie orthogonal to the halo angular momentum.
In both cases, the correlations are small, but detectable.

In this paper, we present an extensive study of the shapes of galaxy
and group-mass halos in a large high resolution \lcdm\ $N$-body simulation.
We study the internal alignments of all of the principal axes
and the angular momentum.
We also study the alignment of all of these quantities with the
large scale structure, and how they are correlated in halos
of different separations.
This work improves upon earlier studies of the internal
structure of halos by using large
high resolution simulations in a currently-favoured \lcdm\ cosmology,
by studying both the alignment of the angular momentum and
the shape, and by using a method that allows us to quantify our errors
and therefore feel confident about the source of any
measured misalignments.
Previous external alignment studies have all been restricted to massive
clusters; we improve upon this significantly by reaching down to
galaxy mass halos, by studying the mass dependence of the correlations,
by studying the alignments of both the principal axes and the angular momenta,
and by including
the oft-neglected intermediate and minor axes.
The structure of the paper is as follows.
In \S~\ref{methodology section} we present the details of the simulation
and describe the method used to measure the principal axes
and angular momentum vectors of the halos.
\S~\ref{shapes results} presents the overall shapes of the
halos and how they change with radius.
We discuss the internal alignment of the principal axes and angular momenta
in \S~\ref{internal alignment results}, while we explore
the alignment of these quantities with external halos
in \S~\ref{external alignment results}.
We discuss what these results imply for the origin
of halo shapes, angular momentum, and warped galaxies
in \S~\ref{discussion section},
and summarize the results in \S~\ref{summary section}.

\section{Methodology}\label{methodology section}

\subsection{The simulation}\label{simulation description}
The simulation used here is the same as the one used in
\citet{bs04-figrot}.
It consists of $512^3$ $N$-body particles
in a periodic box of length $50~\hmpc$, in a low density
flat universe ($\Omega=0.3$, $\Omega_{\Lambda}=0.7$, $h=0.7$,
$\sigma_8 = 0.9$). The particle mass is $7.757\times 10^7~\hmsun$,
and the force softening length is $5~\hkpc$.
Halos were found using the standard friends-of-friends (FOF) algorithm
with a linking length of 0.2 times the mean inter-particle separation.

In order to accurately measure the direction of the principal
axes, a sufficiently large number of particles per halo are required.
The direction of an axis can generally be determined to within
an angle of
\begin{equation}\label{theta err definition}
\theta_{\mathrm{err}} = \frac{1}{2\sqrt{N}} \frac{\sqrt{r}}{1-r}
\mathrm{\quad radians},
\end{equation}
where $N$ is the number of particles used and $r$ is the relevant
axis ratio: $b/a$ for the major axis, $c/b$ for the minor axis,
and $\max(b/a,c/b)$ for the intermediate axis
(see \citealp{bs04-figrot}).
For the purposes of measuring internal alignments, we would like
the angular errors to be less than 10\degr.
This requires on the order of 200 particles.
Since each halo is split up into 6 radial shells, and this accuracy is required
in each shell, each halo should have over 1200 particles.
For convenience, we adopt a cutoff of
$10^{11}~\hmsun$ for the mass of the halo, or 1289 particles.

There are 3869 halos in the sample with masses extending from
$10^{11}~\hmsun$ to $2.8\times 10^{14}~\hmsun$.
451~of the halos have masses in the range $10^{12}$ -- $10^{13}~\hmsun$,
while 62~of the halos
have masses greater than $10^{13}~\hmsun$.

\subsection{Measuring the axes}\label{measuring methodology}

A standard technique to measure halo triaxiality
in simulations is to use an
iterative approach, where the particles are initially chosen
to lie in a sphere or spherical shell, an ellipsoid is fit to
these particles, and particles are chosen for the next iteration
based on the new ellipsoid
\citepalias[e.g.][]{warren-etal92}.
While this works for simulations that have sufficiently low
resolution that overmerging erases substructure,
we find in agreement with \citetalias{js02} that in high
resolution simulations, the presence of substructure prevents
this technique from converging for a large fraction of halos.

\citetalias{js02} have adopted a novel approach aimed at directly
measuring isodensity contours. They assign SPH-like densities
to halo particles, and then measure the principal axes and axis
ratios of particles
with densities near the nominal density of the isodensity contour.
Due to the presence of
substructure, this procedure often picks out disconnected shells
in addition to
the particles that define the isodensity contour of the smooth distribution.
\citetalias{js02} use the FOF algorithm to select the largest
structure that fulfils the density criterion and therefore
eliminate the substructure.

While this algorithm works well for very high-resolution halos,
such as those \citetalias{js02} use to demonstrate the technique
(all of which contain $N > 6\times 10^5$ \citep{js00}),
we encountered difficulties using it on
more moderate resolution halos. In particular, we find
that it is not possible to find an
optimal FOF linking length for eliminating substructure;
if the linking length is too large, many ``contours'' contain
obvious disconnected substructures,
while reducing the linking length sufficiently
to eliminate this problem reduces it to the point where in many halos,
the single ellipsoid corresponding to the smooth distribution is
broken up by the algorithm into several disconnected pieces.
\citetalias{js02}'s algorithm also uses a relatively small number
of particles to determine the shape. The error in the determination
of the direction of the
principal axes of a particle distribution goes as $N^{-1/2}$
(see eq.~[\ref{theta err definition}]).
Measuring the internal alignment of a halo
requires small well-understood errors, and therefore as many particles
as possible.

We take the following approach,
similar to that taken by \citet{frenk-etal88}.
The center of mass is calculated iteratively in spheres centred
on the center of mass of the sphere in the previous iteration,
starting with a sphere containing all of the particles.
The radius of each successive sphere is reduced by 90\%, and
the procedure is iterated 25 times, by which point it has converged.
The particles of each halo are transformed into this center of mass
frame, and the velocities are transformed into the center of velocity frame.
Each halo is then divided up
into six concentric spherical shells with outer radii $R$ 
of 1.0, 0.6, 0.4, 0.25, 0.12,
and 0.06 times the virial radius \rvir.
These radii are chosen to allow easy comparison to the isodensity
contours of \citetalias{js02}.
The outer radius of each shell also forms the inner radius of the
next larger shell, except for
the innermost ``shell'', which
is actually a sphere extending to the halo center.
One would like to use the inertia tensor to measure the principal
axes of the mass distribution. However, the inertia tensor
can be dominated by substructure in the outer part of the shell.
Therefore, we weight particles by $1/r^2$ so that each mass unit
contributes equally regardless of radius \citep{gerhard83}.
Within each shell, we
calculate this reduced moment of inertia tensor
\begin{equation}\label{inertia tensor definition}
\tilde{I}_{ij} = \sum_k \frac{ r_{k,i} r_{k,j} }{r_k^2},
\end{equation}
which we then diagonalize.
The axis ratios $a$, $b$, and $c$
are the square roots of the eigenvalues ($a \ge b \ge c$),
and the eigenvectors give the directions of
the principal axes.
There are no particles in common between radial shells,
so the measurements of the axes at different radii are completely independent.
We also calculate the angular momentum of
the particles in each shell.

To calculate the error in the axis and angular momentum determinations,
we perform a bootstrap analysis of the particles in each radial
shell \citep{heyl-etal94}. If the shell contains $N$ particles,
we resample the shell by randomly selecting $N$ particles from that
set allowing for duplication and determine the axes and angular
momentum from this bootstrap set. We do this 100 times for each
radial shell. The dispersions of these estimates of the
axis ratios, directions of the
axes, magnitude of the angular momentum, and direction of the angular
momentum
around the measured values
are taken formally as the ``$1\sigma$'' error of each
of these quantities.

\begin{figure}
\plotone{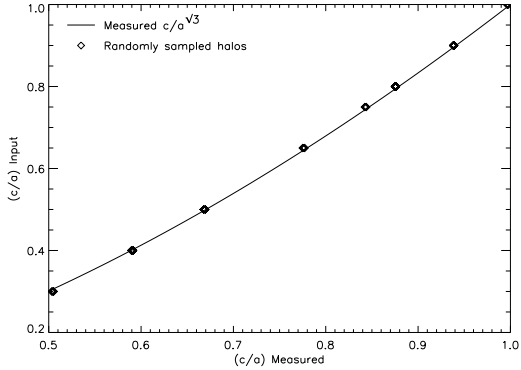}
\caption[Measured versus input $c/a$ axis ratios of randomly-sampled halos]%
{\label{c/a correction figure}%
Measured versus input $c/a$ axis ratio of randomly-sampled
NFW and singular isothermal halos.
At each input axis ratio, there are 12~independent points,
representing the results for each of six radial bins and
two different density profiles, but they are
virtually indistinguishable.
The solid line is the empirical fit of equation~(\ref{c/a correction}).}
\end{figure}

Using spherical shells, rather than the ellipsoids defined by the
isodensity contours as in \citetalias{js02}, does not affect the
orientation of the principal axes. It does, however, bias the derived
axis ratios toward spherical. To calibrate the magnitude of this
bias, we have constructed prolate Poisson-sampled NFW \citep*{nfw96} and
singular isothermal halos with $10^5$ particles each that have known
$c/a$ axis ratios ranging from 0.3 -- 1.
The NFW halos had concentration parameters $c_{\mathrm{vir}}=10$.
In Figure~\ref{c/a correction figure}
we plot the input $c/a$ axis ratio for each
combination of radial bin and
density distribution as a function of the $c/a$ ratio
measured for the halos constructed
using the method described above. The points for different radii
and density distributions are
virtually indistinguishable; the only important
parameter is the input axis ratio. The solid line is an empirical
fit to these points:
\begin{equation}\label{c/a correction}
(c/a)_{\mathrm{true}} = (c/a)_{\mathrm{measured}}^{\sqrt{3}}.
\end{equation}
The relationship for $b/a$ is identical.
The corrected axis ratios $(b/a)_{\mathrm{true}}$ and
$(c/a)_{\mathrm{true}}$ are used throughout the remainder of this paper.

\section{Shapes}\label{shapes results}

\begin{figure}
\plotone{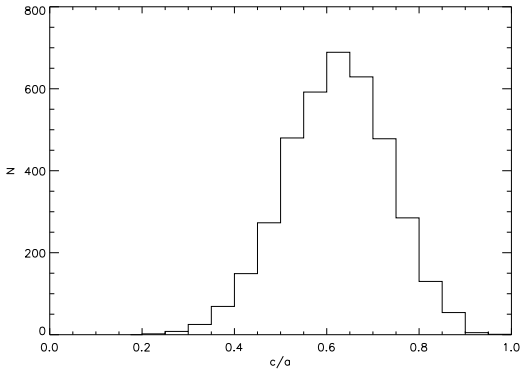}
\caption[Histogram of the minor-to-major axis ratio for each halo]%
{\label{c/a R=0.4 histogram}%
Histogram of the minor-to-major $c/a$ axis ratio for each
halo in the simulation, as measured in the $R=0.4~\rvir$ shell.}
\end{figure}

\begin{figure}
\plotone{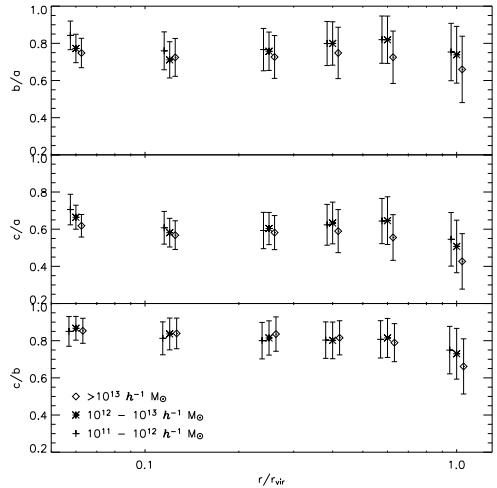}
\caption[Median axis ratios as a function of radius within the halo]%
{\label{b/a and c/a medians}%
Median of the $b/a$ \textit{(top)}, $c/a$ \textit{(middle)},
and $c/b$ \textit{(bottom)}
axis ratios for each radial shell.
Crosses, asterisks, and diamonds
represent the mass ranges $10^{11}$ -- $10^{12}~\hmsun$,
$10^{12}$ -- $10^{13}~\hmsun$, and  $10^{13}$ -- $3 \times 10^{14}~\hmsun$
respectively. Error bars represent the $1\sigma$ width of the
distribution.
The error in the median is typically 0.002 for the crosses,
0.005 for the asterisks, and 0.01 for the diamonds --- smaller than
the sizes of the symbols.
Crosses and diamonds are offset in radius for clarity.}
\end{figure}

Dark matter halos are well approximated by ellipsoids \citepalias{js02}
and are well described by the intermediate-to-major
and minor-to-major axis ratios $b/a$
and $c/a$. Figure~\ref{c/a R=0.4 histogram} is a histogram of the $c/a$
axis ratios as measured in the $R=0.4~\rvir$ shell for all of the halos
in our sample. Unlike \citetalias{js02}, we find that they
are not quite Gaussian, but rather have a tail toward very flattened halos
as seen by \citetalias{ke04},
although the tail is not as extreme as that seen by \citet{bullock02}.
The distribution of $b/a$ and $c/a$ values measured at other radii
have a similar shape.
The lack of very flattened halos in \citetalias{js02} may be a result of
their exclusion of halos deemed to be interacting.
The axis ratios we find are intermediate between the quite flattened
halos found by \citetalias{js02} and the more spherical halos
found by \citetalias{ke04}.

Early studies suggested that
the coldness of the Sgr stream indicates that the dark matter halo
of the Milky Way has $c/a \gtrsim 0.8$
\citep{ibata-etal01,jlm05,martinez-delgado-etal04}.
Only 5\%\ of the halos shown in Figure~\ref{c/a R=0.4 histogram} have
axis ratios so large.
While this is not negligible, it is uncomfortably small,
and forces us into the anti-Copernican situation of living in an
exceptional galaxy.
However, new models of the Sgr stream that are more careful about
matching observations of the body of the dwarf
find that the stars that constitute the stream
were stripped from the body of the satellite too recently to have
had time to undergo differential procession, thereby severely
weakening the constraints on the halo ellipticity \citep{helmi04a,helmi04b};
current models are unable to simultaneously fit the velocities of stars
in the leading stream and the orbital planes of the leading and trailing
streams \citep{ljm05}.
It should also be noted that our simulations do not take into account
the effects of baryonic physics. There is some evidence that baryon
cooling leads to more spherical halos \citep{dubinski94,kazantzidis-etal04}.
Therefore systems in which most of the baryons have cooled, such as
disk galaxies like the Milky Way, may have dark matter shapes that
are more spherical than those presented here.
Observations of external galaxies using a variety of methods
find halo flattenings that range from 0.3 to 0.8, in agreement
with our results \citep{sackett99}.

The radial dependence of the axis ratios
is shown in Figure~\ref{b/a and c/a medians}.
There are three distinct regions of the halo. Over most of the halo,
the axis ratios increase with radius (i.e.~the halos
become more spherical).
Near the virial radius, infalling unvirialized structure
causes the axis ratios to drop. In the central 6\%\ of the virial radius,
the axis ratios rise. However, this is probably an artifact of the
numerical softening \citep{bs04-figrot}.
Further evidence that this is a numerical effect comes from examining
the location of this increase in sphericity for halos of different mass.
The increase occurs at a larger fraction of the virial radius for
low mass halos, i.e.~at a similar physical radius.
This complicated and non-monotonic radial dependence may explain
the discrepancy between studies that have found that flattening
increases with radius \citep{frenk-etal88,cl96,bullock02} and those that
have found that it decreases with radius
(\citealp{dc91}; \citetalias{warren-etal92}; \citetalias{js02};
\citetalias{hbb05}).
The mass dependence of the axis ratios is shown by the
different symbols in Figure~\ref{b/a and c/a medians}.
The highest mass halos
have smaller axis ratios at all radii than the smaller halos
(\citetalias{warren-etal92}; \citealp{bullock02}; \citetalias{js02};
\citetalias{ke04}).
Our halos extend down to masses an order of magnitude smaller 
than any other study; the difference between the most massive halos
(the diamonds) and the galaxy mass halos is more pronounced than
the difference between different masses of galaxy halos.
The high mass halos show particularly
strong flattening near the virial radius.

\begin{figure}
\plotone{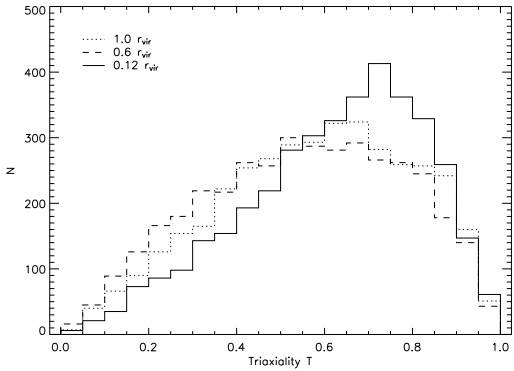}
\caption[Histogram of the triaxiality $T$ of all halos]%
{\label{triaxiality histogram}%
Histogram of the triaxiality $T$ of all halos. The solid,
dashed, and dotted histograms represent the halos measured at
0.12, 0.6, and 1.0~\rvir\ respectively.}
\end{figure}

\begin{figure}
\plotone{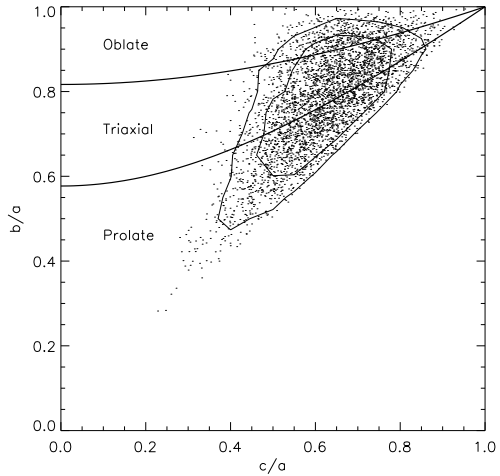}
\caption[Relationship between $b/a$ and $c/a$]%
{\label{b/a vs c/a}%
Intermediate axis ratio $b/a$ as a function of minor axis ratio $c/a$
for all of the halos, measured at $R=0.4~\rvir$. The inner and outer
contours enclose 68\%\ and 90\%\ of the halos respectively.
The thick lines have constant values of the triaxiality parameter $T$.
The separations between the prolate, triaxial, and oblate populations
occur at $T=1/3$ and $2/3$.%
}
\end{figure}

As seen in Figure~\ref{b/a and c/a medians},
the $c/b$ ratio falls steadily with radius, indicating a transition
from prolate figures in the center to oblate figures at large radii
(\citealp{dc91}; \citetalias{warren-etal92}).
Triaxiality can be quantified by the parameter $T$:
\begin{equation}\label{triaxiality T definition}
T = \frac{a^2 - b^2}{a^2 - c^2}
\end{equation}
\citep{fidz91}.
Purely prolate halos have $T=1$ while purely oblate halos have $T=0$.
We have measured $T$ at three radii: $R=1.0~\rvir$, where infalling
material results in substantial flattening, $R=0.6~\rvir$, where
the halos are at their least flattened, and $R=0.12~\rvir$, where
the interior of the halos are at their most flattened.
Histograms of $T$ at these radii are shown in
Figure~\ref{triaxiality histogram}.
The interior regions of halos clearly tend to be prolate (solid histogram).
As the flattening decreases at larger radii, many of the halos
become more oblate (dashed histogram), although still more are prolate
than oblate. Near the virial radius, there is a small shift back toward
prolate shapes (dotted histogram).
Figure~\ref{b/a vs c/a} shows the full relationship between $b/a$ and
$c/a$ for all of the halos in our sample, measured at $R=0.4~\rvir$,
where the values are most typical for the halo as a whole.
The preponderance of prolate and triaxial halos over oblate halos
is clearly seen.

\citetalias{hbb05} examined the evolution of ellipticity of
cluster mass halos at
a wide range of redshifts in a \lcdm\ simulation, and found
that halos of a given mass have higher ellipticity at higher redshift.
In particular, they find
that $\left< e \right> = 0.33 + 0.05z$ for all of their halos, with
slightly stronger evolution ($d\left< e\right> / dz = 0.06 \textrm{--}
0.07$) for their lowest mass halos, which overlap with our high mass halos
(note that these authors define $e\equiv 1-b/a$).
There are two possible interpretations for this result: either
individual halos become more spherical with time, or halos that form
later are intrinsically more spherical and the evolution seen by
\citetalias{hbb05} is due to the growth of individual halos,
resulting in a typically later formation time of halos of a given mass
as time passes.
It is interesting to compare this with the results of
\citet{bs04-figrot}, who studied the evolution of the ellipticity 
of individual halos at $z\approx 0$. \citet{bs04-figrot} find that
individual halos have a mean $d(b/a) / dt = 0.007~h^{-1}~\mathrm{Gyr^{-1}}$,
corresponding to $d\left<e \right> / dz = 0.07$, in good agreement with
\citetalias{hbb05}. We conclude that the shapes of individual halos become more spherical with time.

\begin{figure}
\plotone{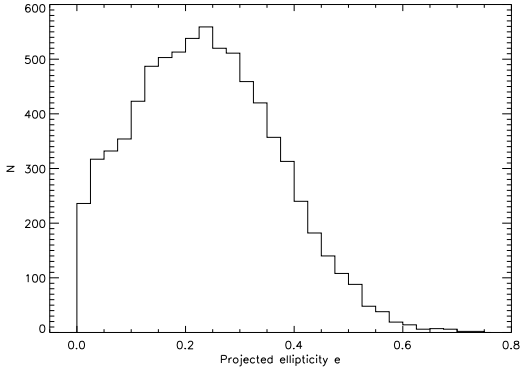}
\caption[Histogram of projected ellipticities]%
{\label{ellipticity histogram}%
Projected ellipticity $e$ of each halo seen from two random orientations,
measured at $R=0.4~\rvir$.}
\end{figure}

Weak lensing measurements have recently begun to probe the
two-dimensional projected
ellipticity of the lensing mass distribution \citep{hyg04}.
For each halo in our sample, we have calculated the
projected axis ratio $q$ using the method of
\citet{stark77} for two random orientations, and calculated the
ellipticity $e$ using equation~(\ref{ellipticity definition}).
A histogram of the results for the $R=0.4~\rvir$ shell
is shown in Figure~\ref{ellipticity histogram}.
The mean and median of the distribution of ellipticities are
$e_{\mathrm{mean}} = 0.24$ and $e_{\mathrm{median}} = 0.23$
respectively, with a $1\sigma$ width of 0.13.
This is consistent with the lower limit of
$\left< e \right> = 0.33_{-0.09}^{+0.07}$ found by
\citet{hyg04} from stacked weak lensing measurements
around galaxies.
It is smaller than the ellipticities found by
\citet{cm00} (note that they quantify ellipticity as
$e = (1-q^2)/(1+q^2)$, in which units our
mean ellipticity is $e=0.27$).
However, most of the halos used in their study
have higher masses than the galactic-mass halos studied here,
and they included relatively poorly-resolved halos which may skew
the results toward higher ellipticities.

\section{Internal alignment}\label{internal alignment results}

\subsection{Principal axes}\label{principal axes internal alignment results}

We compare the alignment of the principal axes within each halo to
see whether the approximation of the halo as a set of concentric
ellipsoids is justified \citepalias{js02}.
In order to determine whether the axes are aligned, the directions
of the axes must be well determined, otherwise apparent  misalignments may be produced owing to  measurement errors.
Therefore, we restrict ourselves
in this section to axes whose bootstrap error is less than 0.2~radians.
The number of halos satisfying this criterion at each radius for each
axis is given in Table~\ref{good measurement table}.

\begin{deluxetable}{lccc}
\tablewidth{0pt}
\tablecaption{Number of halos with axes determined to within 0.2~radians%
\label{good measurement table}}
\tablehead{ \colhead{Radius} & \colhead{Major axis} &
 \colhead{Intermediate axis} & \colhead{Minor axis} \\
\colhead{(\rvir)} & & & \\}
\startdata
0.06 & 904 & 388 & 621\\
0.12 & 2370 & 1005 & 1628\\
0.25 & 2942 & 1762 & 2540\\
0.4 & 2545 & 1488 & 2421\\
0.6 & 2322 & 1363 & 2324\\
1.0 & 2973 & 2231 & 2944\\
\enddata
\end{deluxetable}

To understand the effect of the error on the determination of the 
alignment, consider two axes which
are intrinsically perfectly aligned, but are each measured with an error
of 0.2~radians (note that this is the worst possible case --- the median
error of the sample is 0.1~radians).
Due to the measurement error, these axes will appear to be misaligned
by an angle $\theta_{\mathrm{spurious}}$.
The component of an isotropic error in any particular plane, such as the plane
containing both of the measured axes, is half of the isotropic error,
so we divide the isotropic error of 0.2~radians by two and add
the error of each axis
in quadrature to find the typical $\theta_{\mathrm{spurious}} \approx 0.14$.
Therefore, the cosine of the angle between the two axes,
which is intrinsically $1.0$,
is measured to be $\cos\theta_{\mathrm{spurious}}=0.99$.
If the axes are intrinsically perpendicular, in which case the effect of
$\theta_{\mathrm{spurious}}$ on the direction cosine is maximized,
the error in the direction cosine is $0.14$.
Most halos have well aligned axes (see below), so
the error is negligible.

\begin{figure}
\plotone{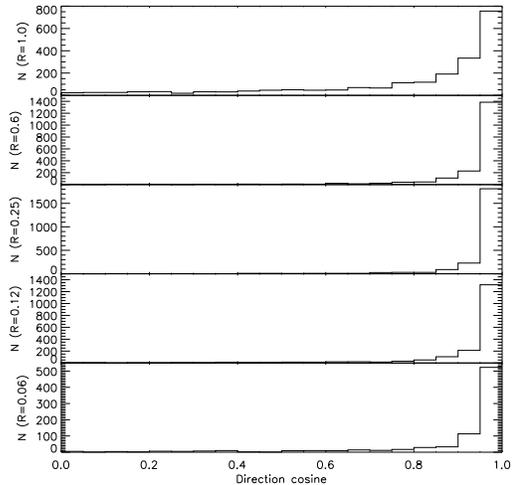}
\caption[Histogram of the internal alignment of the major axes]%
{\label{a dot a internal histogram}%
Histograms of the direction cosine between the major axis of the halo
at $R=0.4~\rvir$ and the major axis at $R=1.0$, 0.6, 0.25, 0.12, and
0.06~\rvir\ (top to bottom).
Due to the symmetry of the axes, this is always positive.
Each histogram contains all halos where the major axes at both
radii are determined to within 0.2~radians.
If the axes were isotropic, this distribution would be uniform.}
\end{figure}

\begin{figure}
\plottwo{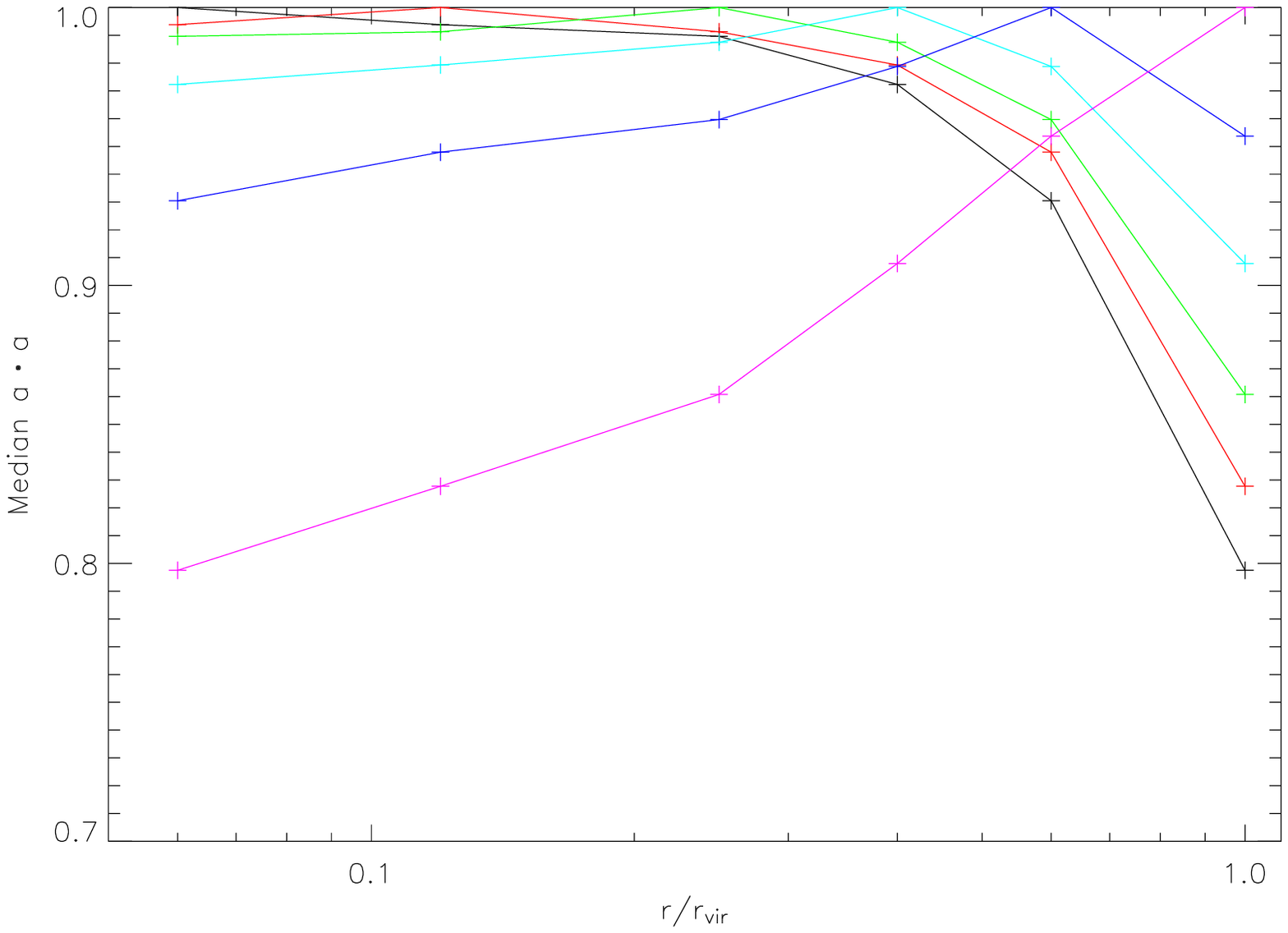}{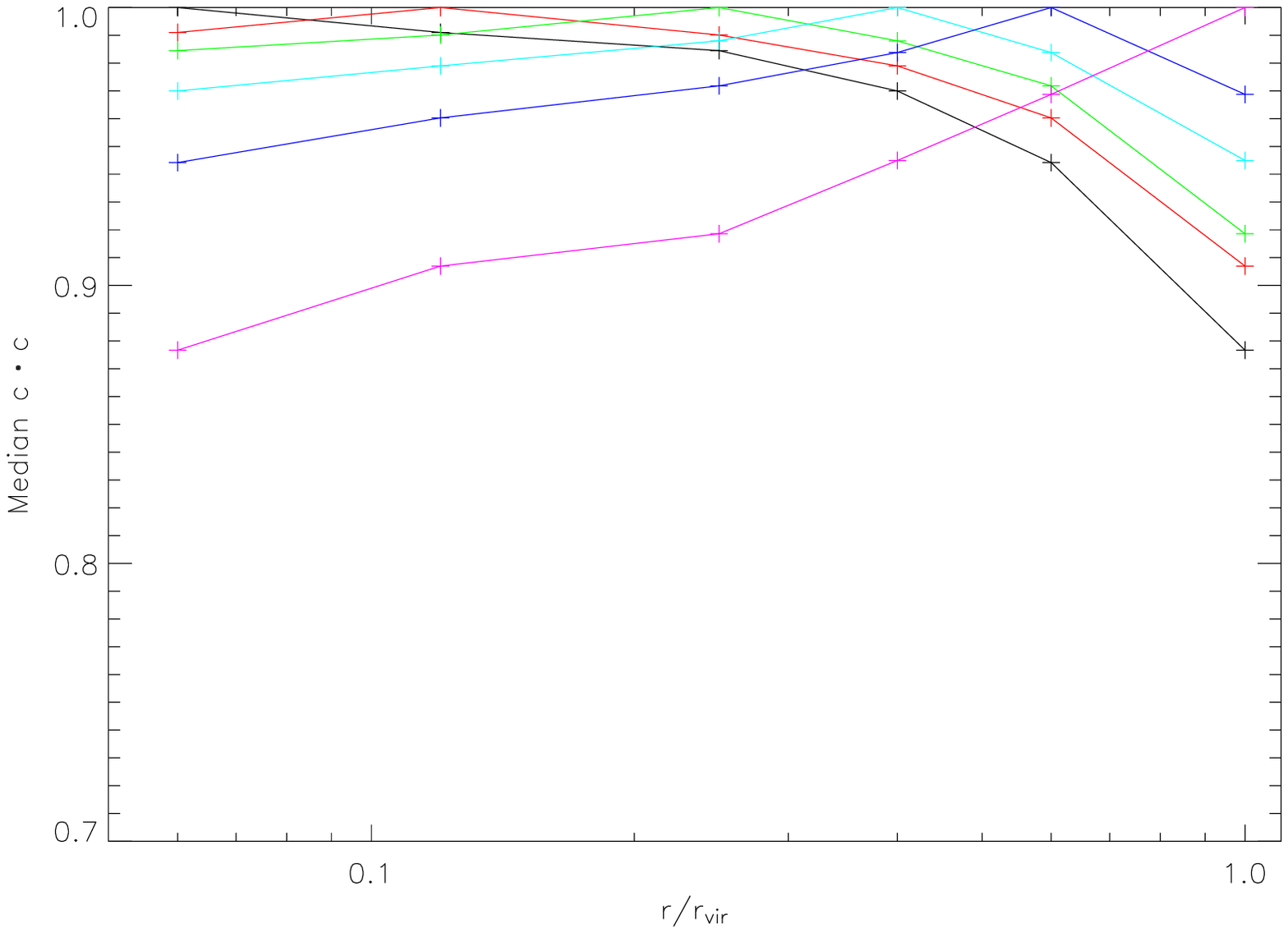}
%\plottwo{f8a-bw.eps}{f8b-bw.eps}
\caption[Median internal alignment of the principal axes as a
function of radius within the halo]%
{\label{internal median axis alignment}%
Median alignment of the major \textit{(a-left)} and minor \textit{(b-right)}
axes at different radii.
The alignment is with respect to the
$R=0.06$ (black/solid), $R=0.12$ (red/dotted), $R=0.25$ (green/short-dashed),
$R=0.4$ (cyan/dot-dashed), $R=0.6$ (blue/dot-dot-dot-dashed),
and $R=1.0~\rvir$ (magenta/long-dashed) shell.
For each pair of radii, only halos where the direction of the
axis is determined to within 0.2~radians at both radii are used.}
\end{figure}

Figure~\ref{a dot a internal histogram} shows histograms of
the alignment between the major axis of
the $R=0.4~\rvir$ shell and the major axis of the
outer (top two panels) and inner (bottom three panels)
regions of the halo. The alignment is very good at all radii.
The relative alignment of the
major and minor axes as a function of radius
is shown in Figure~\ref{internal median axis alignment}.
Each line shows the median alignment
with respect to a different fiducial radius, recognizable as the
radius where the median is exactly unity. The axes are
extremely well aligned within 0.6~\rvir.
Near the virial radius, some halos show
deviations, but there is still usually very good alignment,
especially for the minor axis.
The alignment is better than that seen by \citetalias{js02}.
This confirms their suggestion that many of the halos in which
they measured poor alignment were nearly prolate or oblate.
Such halos have large errors in their direction determination, and
so are not included in our sample.

\begin{figure}
\plotone{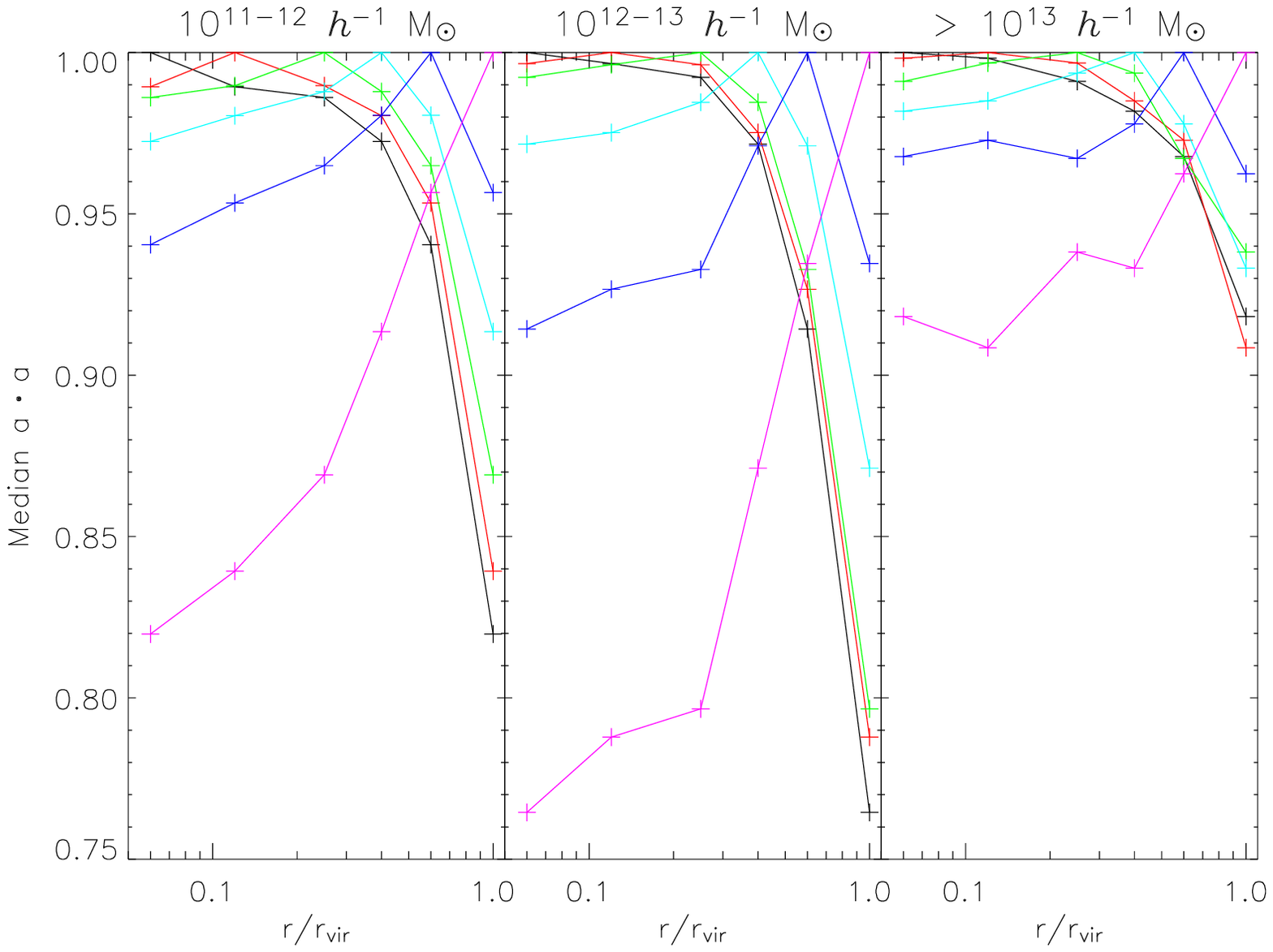}
%\plotone{f9-bw.eps}
\caption[Median internal alignment of the major axes for halos of different mass]%
{\label{a dot a internal vs mass}%
Median alignment of the major axis,
as in Figure~\ref{internal median axis alignment}a,
but only for halos with masses of $10^{11}$ -- $10^{12}~\hmsun$ \textit{(left)},
$10^{12}$ -- $10^{13}~\hmsun$ \textit{(middle)}, and
$10^{13}$ -- $3 \times 10^{14}~\hmsun$ \textit{(right)}.
The alignment is with respect to the
$R=0.06$ (black/solid), $R=0.12$ (red/dotted), $R=0.25$ (green/short-dashed),
$R=0.4$ (cyan/dot-dashed), $R=0.6$ (blue/dot-dot-dot-dashed),
and $R=1.0~\rvir$ (magenta/long-dashed) shell.
For each pair of radii, only halos where the direction of the
axis is determined to within 0.2~radians at both radii are used.}
\end{figure}

Figure~\ref{a dot a internal vs mass} examines the internal alignment
of the major axis as a function of halo mass. The inner 0.4~\rvir\ of
the halos are equally well aligned for halos of all masses. However,
the outer half of the halo
is better aligned with the rest of the halo in high mass halos than
in low mass halos.
\citetalias{js02} saw a similar effect and
suggested that it was because the low mass
halos are intrinsically rounder and therefore have larger errors.
We rule out this explanation, as halos with large errors are not
included in our sample for any mass. Therefore, the stronger alignment
within high mass halos appears to be a real effect.

\subsection{Angular momentum}\label{angular momentum internal alignment results}

\begin{figure}
\plotone{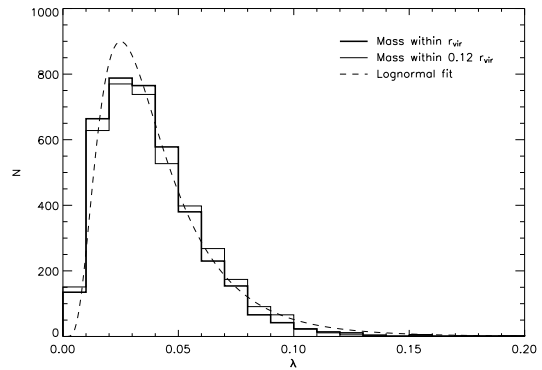}
\caption[Histogram of spin parameters $\lambda$]%
{\label{P(lambda) histogram}%
Histogram of spin parameters $\lambda$ for the halos in our sample.
The thick line shows the distribution when all of the mass in the halo
is used to calculate $\lambda$, while the thin line shows the distribution
when only the central $0.12~\rvir$ of the halo is used.
The dashed line is the lognormal fit to the distribution, where
$\lambda_0=0.035$ and $\sigma=0.58$.}
\end{figure}

Before we examine the orientation of the angular momentum, it is worth
commenting on its magnitude in our simulated
halos. The angular momentum is
usually quantified by the spin parameter $\lambda$, where
\begin{equation}\label{peebles lambda}
\lambda \equiv \frac{J \left| E \right|^{1/2} } { G M^{5/2} }
\end{equation}
\citep{peebles69}.
We use the computationally simpler $\lambda'$ as an estimate for $\lambda$,
where
\begin{equation}\label{lambda prime defn}
\lambda' \equiv \frac{J}{\sqrt{2} M V R}
\end{equation}
\citep{bullock-etal01-angmom}.
In Figure~\ref{P(lambda) histogram}, we plot the distribution of spin
parameters for our halos.
The thick line denotes the distribution we find when we use all of the
mass inside \rvir\ to calculate $\lambda$.
The spin parameters are well described by a lognormal
distribution, shown as the dashed line:
\begin{equation}\label{lognormal lambda}
P(\lambda) = \frac{1}{\lambda \sqrt{2\pi} \sigma} \exp\left(
  - \frac{\ln^2(\lambda/\lambda_0)}{2\sigma^2} \right),
\end{equation}
with $\lambda_0=0.035$ and $\sigma=0.58$, in good agreement with
the results of other simulations \citep{bullock-etal01-angmom}.
The distribution of spin parameters remains essentially unchanged if
we only include mass within the central $0.12~\rvir$ of the halo,
as seen by the thin histogram in Figure~\ref{P(lambda) histogram},
indicating that the angular momentum is distributed evenly throughout
the halo. In the very innermost $0.06~\rvir$, the spin parameters
rise, with the median value of $\lambda$ increasing
from $\lambda_0=0.035$ to $0.047$. However, we do not believe that
this is a physical effect, as the particles in this region
are affected by the numerical force softening, which results in
more tangential orbits and correspondingly larger spin parameters.

\begin{figure}
\plotone{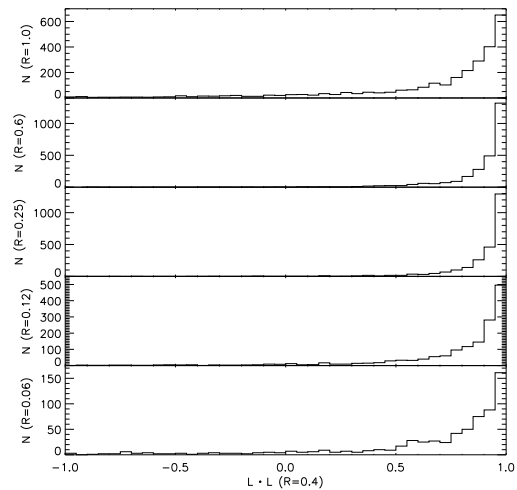}
\caption[Histogram of the internal alignment of the angular momentum]%
{\label{l dot l internal histogram}%
Histograms of the direction cosine between the angular momentum of the halo
at $R=0.4~\rvir$ and the angular momentum at $R=1.0$, 0.6, 0.25, 0.12, and
0.06~\rvir\ (top to bottom).
Each histogram contains
halos whose angular momentum direction is determined to within 0.4~radians
at both radii of the comparison.
If the orientations were isotropic, this distribution would be uniform.}
\end{figure}

\begin{deluxetable}{lc}
\tablewidth{0pt}
\tablecaption{Number of halos with angular momentum directions
determined to within 0.4~radians\label{good L table}}
\tablehead{ \colhead{Radius} & \colhead{Halos}}
\startdata
0.06 & 702\\
0.12 & 1686\\
0.25 & 2820\\
0.4 & 3060\\
0.6 & 3094\\
1.0 & 3229\\
\enddata
\end{deluxetable}

The orientation of the angular momentum cannot be determined as
precisely as the orientation of the principal axes. In order to have
a reasonably large sample of halos, we use
angular momentum vectors whose bootstrap error is less than
0.4~radians. This is twice as large as the limit adopted for
the principal axis directions. The number of halos that satisfy
this criterion at each radius is given in Table~\ref{good L table}.
Following the logic of \S~\ref{principal axes internal alignment results},
the error in the direction cosine of two vectors with errors of
0.4~radians is 0.04 if they are perfectly aligned and 0.28 if they
are perpendicular.
The median errors of the samples are half of these worst-case scenarios.
The alignment is shown to be good (see below), so the effect of the errors
is negligible.

Figure~\ref{l dot l internal histogram} shows
histograms of the relative alignment of
the angular momentum of the $R=0.4~\rvir$\ shell with the outer
regions of the halo (upper two panels) and the inner regions of the halo
(lower three panels).
The alignment is very good at most radii.

\begin{figure}
\plotone{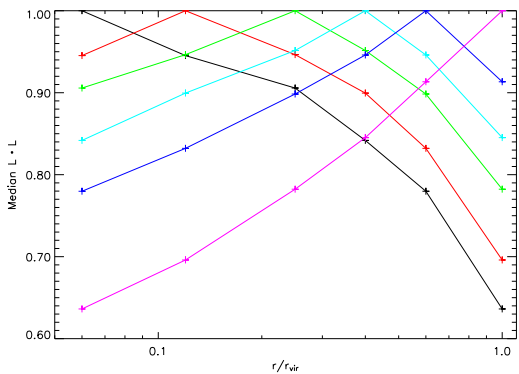}
%\plotone{f12-bw.eps}
\caption[Median internal alignment of the angular momentum as a function
of radius within the halo]%
{\label{internal median angular momentum alignment}%
Median alignment of the angular momentum vector at different radii.
The alignment is with respect to the
$R=0.06$ (black/solid), $R=0.12$ (red/dotted), $R=0.25$ (green/short-dashed),
$R=0.4$ (cyan/dot-dashed), $R=0.6$ (blue/dot-dot-dot-dashed),
and $R=1.0~\rvir$ (magenta/long-dashed) shell.
For each pair of radii, only halos where the direction of the
angular momentum
vector is determined to within 0.4~radians at both radii are used.}
\end{figure}

The relative alignment of the angular momentum as a function of radius
is shown in
Figure~\ref{internal median angular momentum alignment}.
Each line shows the median alignment of the angular momentum with
respect to a different fiducial radius, recognizable as the
radius where the median is exactly unity.
The alignment gets progressively worse as the radii get further
separated; the median cosine between the angular momenta
in the innermost and outermost regions is 0.64.
However, the angular momentum vector at intermediate radius,
such as at 0.4~\rvir, is generally representative of its direction
at all radii.

\begin{figure}
\plotone{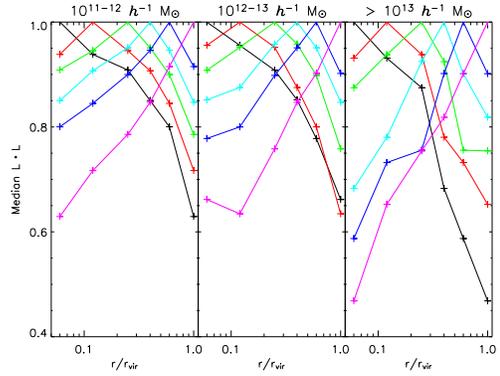}
%\plotone{f13-bw.eps}
\caption[Median internal alignment of the angular momentum, for halos
of different mass]%
{\label{l dot l internal vs mass}%
Median angular momentum alignment
as in Figure~\ref{internal median angular momentum alignment},
but only for halos with masses of $10^{11}$ -- $10^{12}~\hmsun$ \textit{(left)},
$10^{12}$ -- $10^{13}~\hmsun$ \textit{(middle)}, and
$10^{13}$ -- $3 \times 10^{14}~\hmsun$ \textit{(right)}.
The alignment is with respect to the
$R=0.06$ (black/solid), $R=0.12$ (red/dotted), $R=0.25$ (green/short-dashed),
$R=0.4$ (cyan/dot-dashed), $R=0.6$ (blue/dot-dot-dot-dashed),
and $R=1.0~\rvir$ (magenta/long-dashed) shell.
For each pair of radii, only halos where the direction of the
angular momentum
vector is determined to within 0.4~radians at both radii are used.}
\end{figure}

Figure~\ref{l dot l internal vs mass} examines the internal alignment
of the angular momentum as a function of halo mass.
The patterns seen in Figure~\ref{internal median angular momentum alignment}
generally hold for all masses, although the alignment between the very
innermost and outermost regions is slightly worse for the highest
mass halos.

\subsection{Alignment between the angular momentum and the halo shape}\label{l dot axes results}

\begin{figure}
\plotone{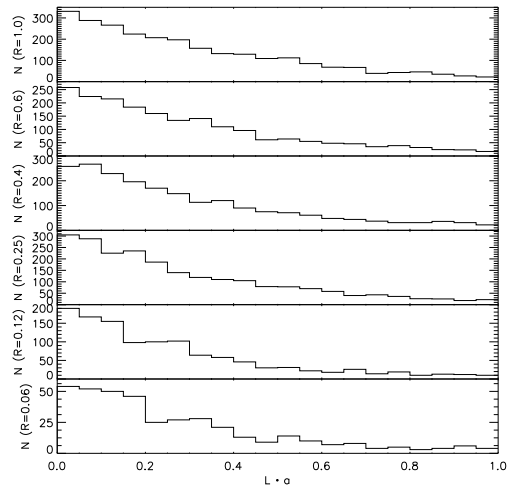}
\caption[Histograms of the alignment of the angular momentum and the
major axis]%
{\label{l dot a internal histogram}%
Histograms of the alignment between the angular momentum vector and the
major axis
of the $R=1.0$, 0.6, 0.4, 0.25, 0.12, and 0.06~\rvir\ (top to bottom) shell
of each halo where the error in the major axis direction is less than
0.2~radians and the error in the angular momentum direction is less
than 0.4~radians at that radius.
Due to the symmetry of the axes, the direction cosine is always positive.
If the orientations were random, this distribution would be uniform.}
\end{figure}

\begin{figure}
\plotone{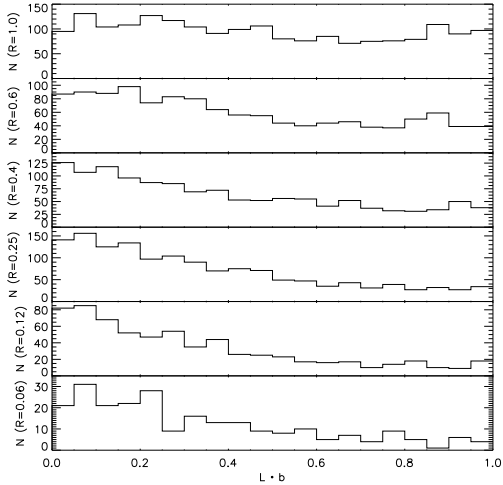}
\caption[Histograms of the alignment of the angular momentum and the
intermediate axis]%
{\label{l dot b internal histogram}%
As in Figure~\ref{l dot a internal histogram} but for
the intermediate axis.}
\end{figure}

\begin{figure}
\plotone{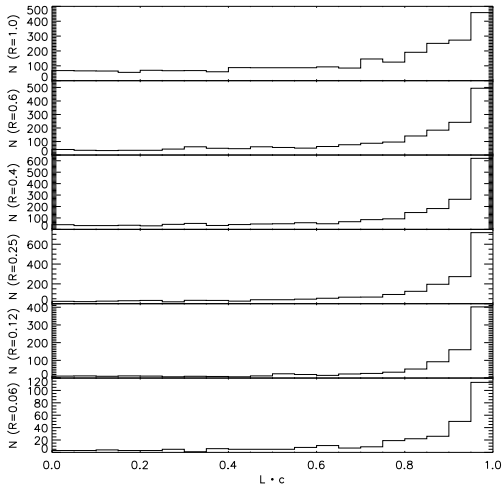}
\caption[Histograms of the alignment of the angular momentum and the
minor axis]%
{\label{l dot c internal histogram}%
As in Figures~\ref{l dot a internal histogram} and 
\ref{l dot b internal histogram}
but for the minor axis.}
\end{figure}

We investigate here the alignment between the angular momentum of a halo at a
given radius and the principal axes of the mass distribution at that radius.
If we compare an angular momentum vector whose error is
0.4~radians with a principal axis whose error is 0.2~radians, the error
in the direction cosine is 0.02 if they are perfectly aligned and 0.22
if they are perpendicular.
The median errors are half of those values. Therefore, in the cases
where the alignment is good, the effect of the errors is negligible.
In the cases where the vectors are perpendicular or isotropic,
we must take the errors into account when drawing conclusions.

Figures~\ref{l dot a internal histogram}, \ref{l dot b internal histogram},
and \ref{l dot c internal histogram} show histograms of the cosine between
the direction of the angular momentum vector and the major, intermediate,
and minor axis respectively. Only those halos with both angular momentum
direction errors of less than 0.4~radians and axis direction errors of less
than 0.2~radians are used. The angular momentum vector tends to be
perpendicular to the major
and intermediate axes, and parallel with the minor axis.
Because of the different effects of the error on parallel and
perpendicular vectors, the tendency of the angular momentum to
be perpendicular to the major axis is as significant as the trend
for it to be parallel to the minor axis,
despite the different appearance of the histograms.
The angular momentum tends to lie perpendicular to the intermediate axis,
but this trend is weaker at larger radii.

These results are consistent with those of \citet{be87}, \citet{dubinski92},
and \citetalias{warren-etal92}.
Of these, only \citetalias{warren-etal92} quantify any change with radius;
they found slightly better alignment at larger radii, in contrast
to the results presented here.
However, both of the radii at which they
performed the comparison were well within the virial radius, well inside
the radii where we see the alignment drop.

\begin{figure}
\plotone{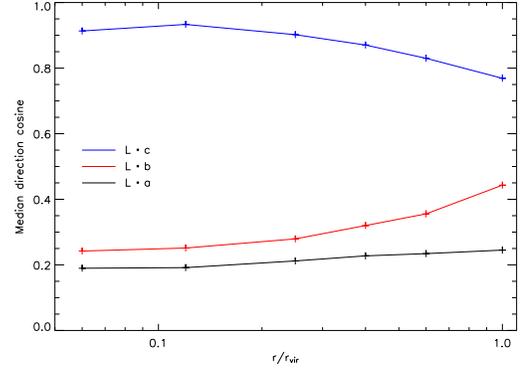}
%\plotone{f17-bw.eps}
\caption[Median alignment of the angular momentum and the
principal axes as a function of radius within the halo]%
{\label{l dot internal axes medians}%
Median alignment between the angular momentum vector and the major (black/solid),
intermediate (red/dotted), and minor (blue/dashed) axis of each halo
as a function of radius within the halo.}
\end{figure}

\begin{figure}
\plotone{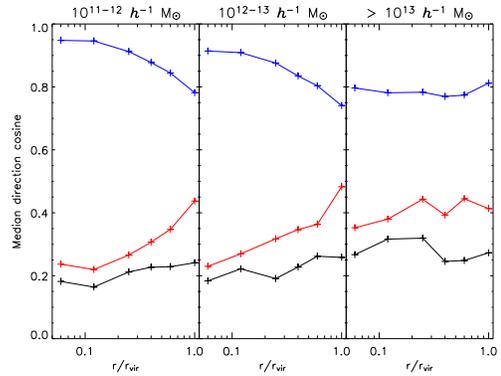}
%\plotone{f18-bw.eps}
\caption[Median alignment of the angular momentum and the principal
axes for halos of different mass]%
{\label{l dot internal axes vs mass}%
As in Figure~\ref{l dot internal axes medians},
but only for halos with masses of $10^{11}$ -- $10^{12}~\hmsun$ \textit{(left)},
$10^{12}$ -- $10^{13}~\hmsun$ \textit{(middle)}, and
$10^{13}$ -- $3 \times 10^{14}~\hmsun$ \textit{(right)}.}
\end{figure}

These relationships are summarized in Figure~\ref{l dot internal axes medians},
which shows the median alignment between the angular momentum vector and
each of the principal axes. The angular momentum tends to lie parallel
with the minor axis and perpendicular to both the major and intermediate
axes. These trends are strongest in the central 0.25~\rvir\ of the
halo, deteriorating slightly in the outer regions.
The median cosine of 0.9 corresponds to a misalignment between the
angular momentum and minor axis of $25\degr$.
Figure~\ref{l dot internal axes vs mass} shows how these trends
depend on the halo mass.
For high mass halos, the alignment is slightly worse
and has less of a dependence on the location within the halo.

\section{External alignment}\label{external alignment results}

\subsection{Introduction}\label{external alignment intro}

In this section, we compare the orientation of the principal axes and
angular momenta of individual halos with the location of mass around them
and the orientation of those properties in surrounding halos.
In \S~\ref{internal alignment results} we demonstrated that
the properties of halos in the $R=0.4~\rvir$ shell are
characteristic of their properties over a large range
of radii.
Therefore, we use the $R=0.4~\rvir$ measurements when
comparing to other halos.

For each halo, the volume is split into 7 radial bins.
The nearest bin spans separations from 0 to 390.625~\hkpc. The outer radii
double for each subsequent bin, while the inner radius is equal to the
outer radius of the interior bin.
The largest bin has an outer radius of 25~\hmpc,
extending to the edge of the periodic box.  
The nominal radius of each bin is the outer radius divided by
$\sqrt{2}$; this corresponds to the geometric mean between its
inner and outer radius for all but the innermost bin.
The inner bin has no formal inner radius,
but in practice is limited by twice the radial extent of the typical
halo, or 250~\hkpc. Halos whose centers of mass are
closer to each other than this are merging,
and are detected as a single object by the group finder.

We follow the nomenclature of \citet{splinter-etal97}, and use
``alignment'' to refer to the tendency of a vector (such as a
principal axis or an angular momentum vector) to point toward or away
from other halos, and ``correlation'' to refer to the tendency of
vectors in different halos to point in the same direction.

\begin{figure}
\plotone{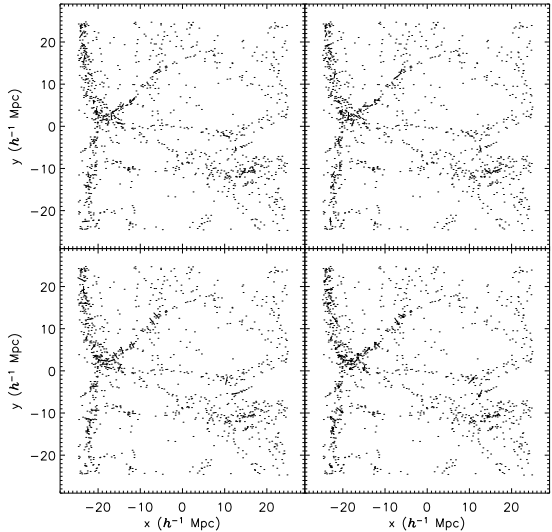}
\caption[Orientation of principal axes and angular momenta
of high mass halos in one quarter of the simulation volume]%
{\label{cluster xyz}%
Projection in $xy$ of the major \textit{(top-left)},
intermediate \textit{(top-right)},
and minor \textit{(bottom-left)} axes,
and unit angular momentum vectors \textit{(bottom-right)}
of high mass ($M > 10^{13}~\hmsun$) halos in a slab
of depth $\Delta z = 12.5~\hmpc$ (one quarter of
the simulation volume). The three-dimensional length of each line
or vector is 2~\hmpc, including the unseen $z$ component.
Dots show the positions of all halos, regardless of mass.}
\end{figure}

A visual impression of how the axes and angular momenta align can be
seen in Figure~\ref{cluster xyz}, which shows the axes and angular momentum
vectors of the high mass halos ($M > 10^{13}~\hmsun$), with the
location of the other halos shown as dots.
Because of the filamentary nature of the large scale structure
\citep[e.g.][]{colberg-etal99},
positive alignment indicates that a quantity tends to point along filaments.

\subsection{Axis alignments}\label{axis alignments results}

We compare here the alignment of the principal axes of the halos with
the location of surrounding structure.
Methods of measuring these alignments
vary in the literature, as does the nomenclature for a given
metric. We adopt an internally-consistent nomenclature
$\xi_{xy}$ and $\xi_{|xy|}$,
which are defined to be the mean value of the direction
cosine between direction $x$ and direction $y$, and the mean of the
absolute value of the direction cosine respectively.
In general, $\xi_{xy}$ is positive if $x$ and $y$ point
in the same direction
and negative if they point in
opposite directions (this
is not applicable if $x$ or $y$ is symmetric, such as
if it is a principal axis),
while $\xi_{|xy|}$ is greater than $1/2$
if $x$ and $y$ lie parallel to each other
and less than $1/2$ if they lie perpendicular to each other.
We note names other
authors use for the same quantities when applicable.
For example, to measure the alignment of the major axis, whose
direction is defined by the unit vector \vhat{a}, with
the large scale structure,
we calculate the alignment $\xi_{|ar|}$:
\begin{equation}\label{a dot r definition}
\xi_{|ar|}(r) \equiv
\left< | \vhat{a} \cdot \vhat{r} | \right>
\equiv
\frac{1}{N} \sum_{i,j} |\vhat{a}_i \cdot \vhat{r}_{ij}|,
\end{equation}
where the sum over $i$ is over all halos in the primary sample,
the sum over $j$ is over all halos in the secondary sample,
$\vhat{r}_{ij}$ is a unit vector in the direction of the displacement
from halo $i$ to halo $j$,
and $N$ is the number of terms in the double sum.
The primary sample consists of all halos whose major axis is
determined to within 0.2~radians,
while the secondary sample consists of all halos.
We define the alignments $\xi_{|br|}$ and $\xi_{|cr|}$
for the intermediate axis \vhat{b}\ and
the minor axis \vhat{c}\ similarly.
Note that the primary samples used to define $\xi_{|ar|}$,
$\xi_{|br|}$, and $\xi_{|cr|}$ are not identical, as the
set of halos with good major axis determinations is deficient
in very oblate halos, the set of halos with good minor axis determinations
is deficient in very prolate halos, and the set of halos with good
intermediate axis determinations is deficient in both very prolate
and very oblate halos.

\begin{figure}
\plotone{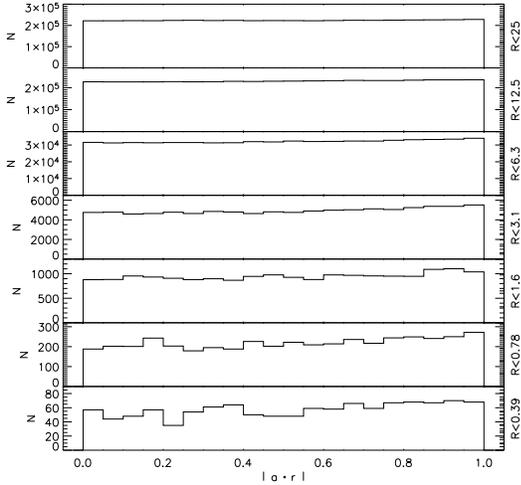}
\caption[Histograms of the alignment of the major axes of halos
and the location of surrounding structure]%
{\label{a dot r external histogram}%
Histograms of the direction cosine $| \vhat{a} \cdot \vhat{r} |$
between the major axes of halos in the primary sample and the
displacement from each primary halo to all surrounding halos,
binned by radial separation between the centers of the halos.
The direction cosine is always positive due to the symmetry of the axes.
If the axes were randomly oriented, the distributions would be uniform.
The radial bins consist of halos separated by (top to bottom):
12.5 -- 25, 6.25 -- 12.5, 3.125 -- 6.25, 1.5625 -- 3.125, 
0.78125 -- 1.5625, 0.390625 -- 0.78125, and 0 -- 0.390625~\hmpc\ respectively.}
\end{figure}

Figure~\ref{a dot r external histogram} shows histograms of the
distribution of direction cosines $| \vhat{a} \cdot \vhat{r} |$ 
for halos at a variety of separations.
If the axes were randomly oriented, the distributions would be uniform.
The distribution is mostly isotropic, but there is an excess
of halos with $|\vhat{a}\cdot\vhat{r}| > 1/2$, ie. halos whose
major axes lie parallel to the filaments.
We quantify this by calculating the mean, $\xi_{|ar|}$
(note that this is similar but not identical to the quantity $w(r)$
used by \citetalias{ke04}).

\begin{figure}
\plotone{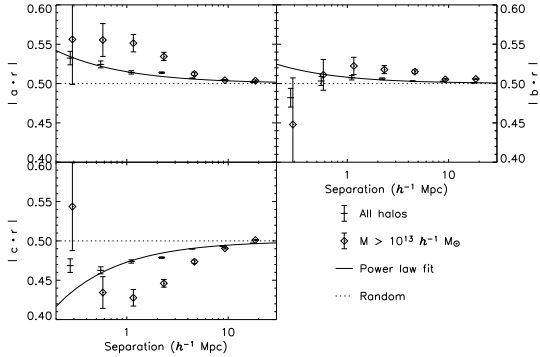}
\caption[Mean alignment of the principal axes of
halos with the surrounding structure]%
{\label{a dot r external}%
Mean alignment of the principal axes of halos in the primary sample
with the displacement from the primary halos to all surrounding halos,
as a function of distance between the halos.
The alignment of the major axis ($\xi_{|ar|}$), intermediate axis
($\xi_{|br|}$), and minor axis ($\xi_{|cr|}$) is plotted in the
top-left, top-right, and bottom-left panels respectively.
Only halos where the relevant axis is determined to within 0.2~radians
are used for the primary sample.
The different symbols are for primary samples consisting of
all such halos (crosses)
or of only those with masses greater than $10^{13}~\hmsun$ (diamonds).
The diamonds are shifted slightly to the right for clarity.
Error bars represent the $1\sigma$ Poisson sampling error in the mean.
The solid lines are the power law fits for the full samples,
and are given by equations~(\ref{a dot r power law fit}), 
(\ref{b dot r power law fit}), and (\ref{c dot r power law fit}).
The dotted line is the value expected for random orientations.}
\end{figure}

The upper-left panel of
Figure~\ref{a dot r external} shows $\xi_{|ar|}$ for the halos.
We find that the alignment is well fit by a power law
over a wide range of separations.
The fit is shown as the solid line in
Figure~\ref{a dot r external}, and is given by
\begin{equation}\label{a dot r power law fit}
\xi_{|ar|}(r) = \frac{1}{2} + m_1 r^\alpha,
\end{equation}
where $r$ is the separation in units of \hmpc, the alignment at
1~\hmpc\ is $m_1=0.015$, and the slope is $\alpha=-0.64$.
\citetalias{faltenbacher-etal02} find even stronger alignment for their
cluster-mass halos, as do \citetalias{ke04} and \citetalias{hbb05}
(except at the very
smallest separations, which is at the spatial limit of their simulations
and the point where halos begin to overlap).
This may be because all of these authors
analyze simulations performed in much larger boxes,
which contain power on longer wavelengths than exist in our
smaller simulation volume (which also allows them to measure
alignments at separations far exceeding our entire box length).
Alternatively, 
the stronger alignment they detect may be a consequence of the higher mass of
their halos.
To see whether the halo mass is important, we
recalculate $\xi_{|ar|}$, restricting the primary
sample to those halos with masses greater than
$10^{13}~\hmsun$.
This is the tendency for the major axes of group or cluster mass
halos to point along the filaments.
The results are shown as the diamonds in Figure~\ref{a dot r external}.
The group and cluster mass halos are much more strongly aligned than the
full sample (which is dominated by galaxy mass halos).
The cluster alignment $\xi_{|ar|}$ is constant out to 3~\hmpc, after
which it drops until it agrees with the results of the full sample
by 6~\hmpc.
\citetalias{hbb05} also find a strong mass dependence;
however, the alignments of their lowest mass sample are still larger
than the alignments of our highest mass sample, indicating that
the box size also plays a role.

The intermediate axes are also aligned with the filaments, as seen
in the upper-right panel of Figure~\ref{a dot r external}.
While the innermost bins show no alignment,
there is clear evidence of alignment at separations of 1~\hmpc\ and beyond.
The solid line is a power law fit to the outer 5 points,
i.e.~for $r>781~\hkpc$, and is given by
\begin{equation}\label{b dot r power law fit}
\xi_{|br|} = \frac{1}{2} + m_1 r^\alpha,
\end{equation}
where $m_1=0.008$ and $\alpha=-0.7$.
The high mass halos again show constant alignment out to several
Mpc.

As both the major and intermediate axes tend to point along filaments,
the minor axis must tend to lie perpendicular to the filaments.
This trend can be clearly seen
in the bottom-left panels of Figure~\ref{cluster xyz}
as sequences of parallel lines crossing the filaments.
The lower panel of
Figure~\ref{a dot r external} shows $\xi_{|cr|}$.
The solid line is a power law fit for the full sample excluding
the innermost bin,
i.e.~for $r>391~\hkpc$, and is given by
\begin{equation}\label{c dot r power law fit}
\xi_{|cr|}(r) = \frac{1}{2} + m_1 r^\alpha,
\end{equation}
where the alignment at 1~\hmpc\ is $m_1=-0.027$ and $\alpha=-0.7$.
The minor axes of group and cluster mass halos show even stronger alignment
than do the galaxy mass halos,
as seen by the diamonds in Figure~\ref{a dot r external}.

\subsection{Axis correlations}\label{axis correlations results}

We compare here the tendency for the principal axes of neighbouring
halos to point in the same direction. The procedure we use is
completely analogous to that used to calculate the alignments
in \S~\ref{axis alignments results}.
The correlation between the major axes is defined as
\begin{equation}\label{xi_aa definition}
\xi_{|aa|}(r) \equiv
\left< | \vhat{a} \cdot \vhat{a} | \right> \equiv
\frac{1}{N} \sum_{i,j} |\vhat{a}_i \cdot \vhat{a}_j|,
\end{equation}
where the sum is over all unique pairs $(i,j)$ because the
measurement is symmetric with respect to each pair of halos.
Only halos whose major axes are determined to within 0.2~radians
are used for both the primary and secondary samples.

\begin{figure}
\plotone{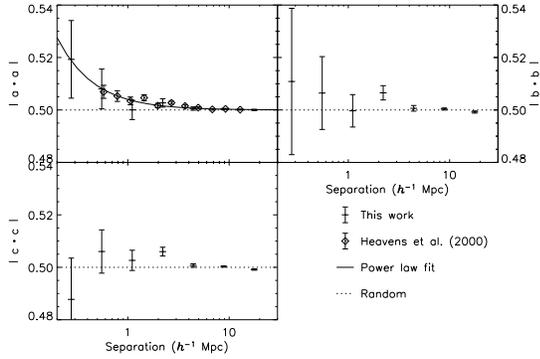}
\caption[Mean correlation of the principal axes of 
neighbouring halos]%
{\label{aa bb cc external all}%
Mean correlation
of the principal axes of halos for which the
direction of the major axes is determined to
within 0.2~radians for both halos.
The correlations of the major axes ($\xi_{|aa|}$), intermediate
axes ($\xi_{|bb|}$), and minor axes ($\xi_{|cc|}$) are plotted
in the upper-left, upper-right, and bottom-left panels respectively.
The solid line is the power law fit for the
major axis correlation, and is given by
equation~(\ref{a dot a power law fit}).
The dotted line is the expected value for random orientations.
Error bars represent the $1\sigma$ Poisson sampling error in the mean.
Diamonds represent the results of \citetalias{hrh00}.}
\end{figure}

The top-left panel of Figure~\ref{aa bb cc external all}
shows the mean correlation $\xi_{|aa|}$ of the major axes
as a function of the halo separation (this is identical
to the quantity defined as $u(r)$ in \citetalias{ke04}).
Although almost all bins are individually consistent with
isotropy, they all lie above 0.5, and taken together are evidence
that the directions of the major axes are correlated.
We fit a power law and find
\begin{equation}\label{a dot a power law fit}
\xi_{|aa|}(r) = \frac{1}{2} + m_1 r^\alpha,
\end{equation}
where $r$ is the separation in units of \hmpc, the correlation at
1~\hmpc\ is $m_1=0.004$, and the slope is $\alpha=-1.2$,
although the large errors introduce considerable uncertainties in these values.
The correlations found by \citetalias{faltenbacher-etal02},
\citetalias{ke04} and \citetalias{hbb05}
are considerably stronger than those found here.
Comparison with the results of \citet{cm00} and \citetalias{hrh00} is
more difficult because these authors measure
projected ellipticity correlations
rather than axis correlations.
The ellipticity correlations are diluted with respect to the axis
correlations, and therefore we should consider their results as
lower limits.
The results of \citetalias{hrh00} are plotted as diamonds
in Figure~\ref{a dot a power law fit},
where we note that
\begin{equation}\label{xi vs e11 mapping}
\xi_{|aa|} = \frac{1}{2} + \frac{\sqrt{3}}{2}\left( \left< e_1 e_1\right>
  + \left< e_2 e_2\right> \right).
\end{equation}
The agreement is excellent. However, because their results
should be considered as lower limits, and \citet{cm00} found
even larger ellipticity correlations in the same simulation,
it appears that the shapes of the halos studied by these authors
also correlate more strongly than the halos in our simulation.
This is likely because our sample is dominated by galaxy mass halos,
while all of the previous studies have analyzed only cluster mass halos.
We have demonstrated in \S~\ref{axis alignments results} that
the higher mass halos are more strongly aligned with the large
scale structure; it is therefore likely that they also show
stronger correlations with each other.
To test this, we have recalculated $\xi_{|aa|}$ using only halos with masses
greater than $10^{13}~\hmsun$ to see if the behaviour of high
mass halos differs from those of lower mass,
but due to the small number of
very high mass halos in our sample,
the errors are too large to draw any conclusions.

The mean correlations of the
intermediate and minor axes of the halos, $\xi_{|bb|}$ and $\xi_{|cc|}$,
are plotted in the upper-right and lower-left panels
of Figure~\ref{aa bb cc external all}.
The error bars are too large to robustly detect any correlation,
though the lower panel shows a suggestive preponderance
of bins with $\xi_{|cc|} > 0.5$. A larger sample of halos
simulated at equally high spatial resolution is required to confirm if
this is real.

\subsection{Angular momentum}

\begin{figure}
\plotone{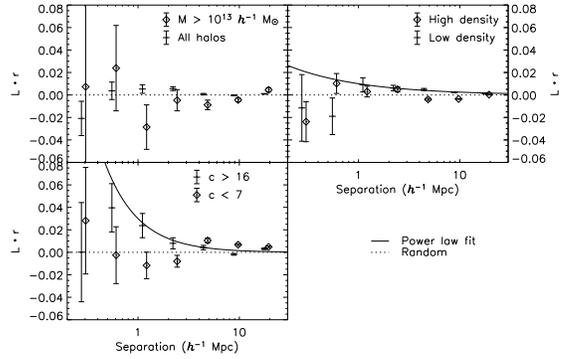}
\caption[Mean vector alignment $\xi_{Lr}$ of the angular momenta
of halos and the location of surrounding structure]%
{\label{l dot r external all}%
Mean vector alignment $\xi_{Lr}$ between the angular momentum
vector of a halo and the location of surrounding halos,
as a function of radial separation.
Error bars represent the $1\sigma$ Poisson sampling error in the mean.
The dotted line is the expected value for random orientations.
Diamonds are shifted to the right for clarity.
\textit{(Top-left):} The alignment of all halos (crosses), and
of only those halos with masses greater than $10^{13}~\hmsun$ (diamonds).
\textit{(Top-right):} The alignment of halos that have
3 or fewer neighbouring halos within 2~\hmpc\ (crosses) compared to
those halos with 4 or more neighbouring halos within 2~\hmpc\ (diamonds).
The solid line is the power law fit for the low density sample
at $r > 781~\hkpc$, and is given by
equation~(\ref{l dot r power law fit})
with $m_1=0.01$ and $\alpha=-0.6$.
\textit{(Bottom-left):} The alignment of halos with concentration
parameter $c_{\mathrm{vir}} > 16$ (crosses) compared to those with
NFW concentration parameter $c_{\mathrm{vir}} < 7$ (diamonds).
The solid line is the power law fit for the high concentration
sample at $r > 391~\hkpc$, and is given by
equation~(\ref{l dot r power law fit})
with $m_1=0.03$ and $\alpha=-1.3$.}
\end{figure}

We investigate the tendency for the angular momentum to point toward
or away from nearby halos, for it to lie parallel or perpendicular
to the filamentary structure, and for the angular momentum vectors
of neighbouring halos to point in the same direction.
To measure the tendency
of halo angular momenta to point toward or away from other halos,
we calculate the alignment
\begin{equation}\label{l dot r definition}
\xi_{Lr}(r) \equiv
\left< \vhat{L} \cdot \vhat{r} \right> \equiv
\frac{1}{N} \sum_{i,j} \vhat{L}_i \cdot \vhat{r}_{ij},
\end{equation}
where $\vhat{L}_i$ is a unit vector in the direction of the angular
momentum vector for halo $i$.
The primary sample consists of all
halos where the direction of the angular momentum is determined to within
0.4~radians, while the secondary sample consists of all halos.
We show the results in the top-left panel of
Figure~\ref{l dot r external all}.
There may be a weak tendency for
the angular momentum to point toward local density enhancements
on scales of 0.5 -- 3~\hmpc, but the size of the error bars makes
such a result tentative. When we restrict the sample to those halos
with masses greater than $10^{13}~\hmsun$, we find no statistical
deviation from random.

\citet{ko92} and \citet{nas04} have
found that galaxies within the local supercluster have
their spins pointing within the supercluster plane, while those at least
2~\hmpc\ from the plane have spins that point toward
or away from the plane. To see
if the behaviour of simulated halos in low and high density regions differs,
we have split the sample into those halos that have 3 or fewer neighbouring halos
within 2~\hmpc\ (the low density sample), and those that have 4 or more
neighbours within 2~\hmpc\ (the high density sample).
There are 2155 and 1714 halos in the low and high density samples
respectively. In the top-right panel of Figure~\ref{l dot r external all}
we have plotted $\xi_{Lr}$ for these two samples. 
By construction, there are very few pairs
(and therefore large error bars)
at small separations in the low density sample,
but beyond 1~\hmpc\ there is
a clear detection of positive alignment in this sample.
The alignment is well
fit by a power law of the form
\begin{equation}\label{l dot r power law fit}
\xi_{Lr} = m_1 r^{\alpha},
\end{equation}
where $r$ is the separation in units of \hmpc, the correlation at
1~\hmpc\ is $m_1=0.01$, and the slope is $\alpha=-0.6$.
The high density sample shows no coherent tendency for the
angular momentum vectors to point toward or away from
density enhancements.
The low and high density samples do not show significant
deviations from the full sample for any of the other
statistics studied.

In order to see how the internal structure of the halo
affects its alignment properties, we have split the sample
into those halos with particularly high NFW \citep{nfw96} concentration
parameters ($c_{\mathrm{vir}} > 16$) and those with
particularly low concentration parameters ($c_{\mathrm{vir}} < 7$).
As demonstrated by the
bottom-left panel of Figure~\ref{l dot r external all}, there is a clear
signal of positive alignment in the high-concentration sample, and none
in the low-concentration sample. The alignment
of the high-concentration sample is well fit by
a power law of the form of equation (\ref{l dot r power law fit}),
with $m_1=0.03$ and $\alpha=-1.3$.
Because halo concentration anti-correlates with
mass \citep{bullock-etal01-profiles}, our high-mass sample is
deficient in high-concentration halos. Therefore, the tendency for
low mass halos to show positive $\xi_{Lr}$ while high mass halos
show no such a tendency is probably a direct consequence
of the correlation between $\xi_{Lr}$ and $c_{\mathrm{vir}}$.

\begin{figure}
\plotone{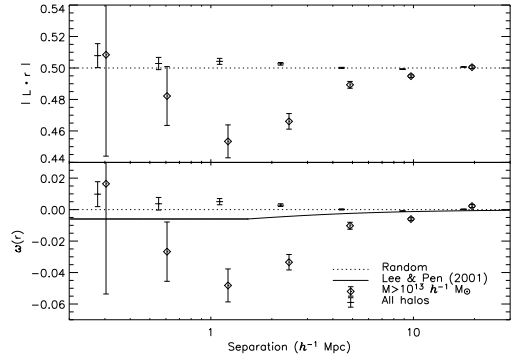}
\caption[Mean alignment $\xi_{|Lr|}$ and $\omega(r)$
of the angular momenta of halos
and the location of surrounding structure]%
{\label{l dot r abs}%
Mean alignment $\xi_{|Lr|}$ \textit{(top)} and $\omega(r)$
\textit{(bottom)}
between the angular momentum vector of primary halos and the location
of all surrounding halos.
The different symbols are for primary samples consisting of
all such halos (crosses),
or of only those with masses greater than $10^{13}~\hmsun$ (diamonds).
The diamonds are shifted slightly to the right for clarity.
Error bars represent the $1\sigma$ Poisson sampling error in the mean.
The dotted line is the expected value for random orientations.
The solid line is the prediction from linear tidal torque theory
\citep{leepen01}.}
\end{figure}

The tendency for the angular momentum vectors to
lie parallel versus perpendicular to the filaments is measured by
\begin{equation}\label{l dot r abs definition}
\xi_{|Lr|}(r) \equiv
\left< | \vhat{L} \cdot \vhat{r} | \right> \equiv
\frac{1}{N} \sum_{i,j} | \vhat{L}_i \cdot \vhat{r}_{ij} |,
\end{equation}
defined as $L_{\parallel}$ in \citet{hn01}, or
\begin{equation}\label{omega definition}
\omega(r) \equiv \left< | \vhat{L} \cdot \vhat{r} |^2 \right> - \frac{1}{3}
\equiv \frac{1}{N} \sum_{i,j} |\vhat{L}_i \cdot \vhat{r}_{ij}|^2 - \frac{1}{3},
\end{equation}
as used by \citet{leepen01}.
We plot these quantities in Figure~\ref{l dot r abs}.
The solid line in the bottom panel shows the prediction
of \citet{leepen01} from linear tidal torque theory.
On scales less than 3~\hmpc, the angular momentum tends
to lie parallel to the filaments. The values are
consistent with those found by \citet{hn01}.
However, both \citet{leepen01} and \citetalias{faltenbacher-etal02}
find that the angular momenta of halos tend to lie perpendicular
to the filaments.
This discrepancy may lie in the
different mass ranges probed. The halos of \citet{hn01}
cover a very similar mass range to those in this work,
while the sample of \citetalias{faltenbacher-etal02} consists entirely
of cluster mass halos (their \textit{smallest} halo has a mass of
$1.4 \times 10^{14}~\hmsun$, nearly the mass of our
\textit{largest} halo).
We have recalculated $\xi_{|Lr|}$ and $\omega(r)$ using only
halos with masses greater than $10^{13}~\hmsun$ in the primary
sample and plotted them as diamonds in Figure~\ref{l dot r abs}.
The behaviour of the high mass halos is radically different; the angular
momenta of groups and clusters tend to point perpendicular to the filaments.

\begin{figure}
\plotone{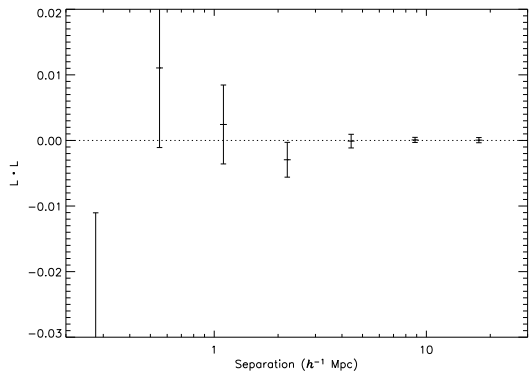}
\caption[Mean vector correlation $\xi_{LL}$ of the angular momenta
of neighbouring halos]%
{\label{l dot l external all}%
Mean correlation $\xi_{LL}$
of the angular momentum vectors of halos as a function
of their separation.
Error bars represent the $1\sigma$ Poisson sampling error in the mean.
The dotted line is the expected value for random orientations.}
\end{figure}

\begin{figure}
\plotone{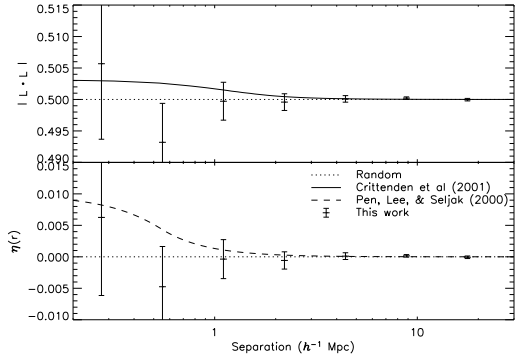}
\caption[Mean correlations $\xi_{|LL|}$ and $\eta(r)$ of the angular momenta
of neighbouring halos]%
{\label{l dot l abs all}%
Mean correlation $\xi_{|LL|}$ \textit{(top)}
and $\eta(r)$ \textit{(bottom)}
between the angular momentum vectors of halos
as a function of their separation.
Error bars represent the $1\sigma$ Poisson sampling error in the mean.
The solid line shows the analytic result of \citet{cnpt01},
where we set the correlation parameter between the shear and inertia
tensors $a=0.55$, while the solid line is the linear tidal torque
prediction of \citet{pls00}.
The dotted line is the expected value for random orientations.}
\end{figure}

The correlation between the directions of the halo
angular momenta is defined as
\begin{equation}\label{l dot l definition}
\xi_{LL} \equiv
\left< \vhat{L} \cdot \vhat{L} \right> \equiv
\frac{1}{N} \sum_{i,j} \vhat{L}_i \cdot \vhat{L}_j,
\end{equation}
where the sum is over all unique pairs $(i,j)$. This is equivalent
to the quantity defined is $\eta(r)$ in \citet{pdh02a}.
As seen in Figure~\ref{l dot l external all}, we detect no
deviations from random. We also measure the tendency for halo
angular momentum vectors to be either parallel or perpendicular
to each other. We define two quantities to measure this,
\begin{equation}\label{l dot l abs definition}
\xi_{|LL|} \equiv
\left< | \vhat{L} \cdot \vhat{L} | \right> \equiv
\frac{1}{N} \sum_{i,j} | \vhat{L}_i \cdot \vhat{L}_j |,
\end{equation}
which is the same as the quantity defined as $\mu(r)$ in \citet{hn01}, and
\begin{equation}\label{l dot l eta definition}
\eta(r) \equiv \left< | \vhat{L} \cdot \vhat{L} |^2 \right> - \frac{1}{3}
\equiv \frac{1}{N} \sum_{i,j} | \vhat{L}_i \cdot \vhat{L}_j |^2 - \frac{1}{3},
\end{equation}
which is the same as \citet{pls00}'s $\eta(r)$ and
\citet{pdh02a}'s $\eta_2(r)$.
The quantities $\xi_{|LL|}(r)$ and $\eta(r)$ are shown in
Figure~\ref{l dot l abs all}.
We detect no deviations from isotropy. 
For reference, we also plot the linear predictions of
\citet{cnpt01} (where we set the correlation parameter between
the shear and inertia tensors $a=0.55$ so that the linear results
also match the $N$-body results of \citetalias{hrh00})
and \citet{pls00}.
Due to the size of the error bars, our results are consistent
with the
non-detections and weak correlations found by other
authors (\citetalias{hrh00}; \citealp{pls00,cnpt01,hn01,pdh02a};
\citetalias{faltenbacher-etal02}).
The number of halos in our sample is too small to detect
angular momentum correlations in any particular subsample.

\section{Discussion}\label{discussion section}

\subsection{Internal alignment and galactic warps}

The internal alignment of the angular momentum provides several clues
to the forces a galaxy undergoes during its evolution.
In particular, we note that, as demonstrated in 
Figure~\ref{l dot internal axes medians}, the angular momentum
and the minor axis of a halo are typically misaligned by $\sim 25\degr$,
even in the central regions of the halo where a galactic disk would lie.
Therefore, if the angular momentum of the dark matter and baryons
are aligned, and the disk lies perpendicular to its angular momentum,
then the disk and halo will be typically misaligned by $\sim 25\degr$.
Such misalignment could be partly responsible for warped galactic disks
\citep{ds83,toomre83,bailin-phd}. However, simulations of live disks
in live halos demonstrate that misaligned disk-halo systems quickly reach
an unwarped realigned equilibrium \citep{bjd98}.

However,
if the orientation of the disk and/or the orientation of the halo changes,
then such an equilibrium may never be established. \citet{qb92} suggested
that the infalling angular momentum may be uncorrelated with the orientation
of the central angular momentum.
Figure~\ref{internal median angular momentum alignment} demonstrates
that this is true in our halos; while the angular momentum at intermediate
radius is generally representative of its orientation at all radii,
the angular momentum of the
innermost and outermost regions are nearly uncorrelated. Therefore,
as the disk accretes angular momentum, its orientation changes.
This reorientation of the disk may itself cause warps
\citep{ob89,lopez-corredoira-etal02a}, or simply prevent the disk and
halo from reaching an aligned equilibrium. Similarly, \citet{bs04-figrot}
found that over 90\%\ of dark matter halos show figure rotation; this
reorientation of the halo also acts to maintain misalignment
between the disk and halo.
Therefore, it appears that misalignment between the halo and disk
is common, and perhaps responsible for galactic warps.

\subsection{External alignment}

The alignment of the halo axes
with the large scale structure seen in Figure~\ref{a dot r external}
provides several clues to the
origin of halo orientations. For instance, the relative alignment
of the major, intermediate, and minor axes is intriguing.
The relative strength of the alignment is well
described by $m_1$, the value of the power law fit at
a separation of 1~\hmpc. This is 0.015, 0.008, and -0.027 for the
major, intermediate, and minor axes respectively.
While all previous authors have neglected the minor axis, we find
that its alignment perpendicular to filaments is stronger than
the alignment of the major axis along filaments.
Therefore it is the minor axis,
not the major axis,
that is most influenced by the presence of surrounding material.
This also explains how both the major and intermediate axes can be
positively aligned with the filament: if the minor axis of a halo
lies perpendicular to the filament, then both the major and intermediate
axes are constrained to lie within a plane that contains the
filament. Therefore, they are both more likely to point along the
filament than a randomly-oriented three-dimensional axis, and therefore
both show positive alignment
(for example, if all of the minor axes were perfectly perpendicular to
the filament and there were no difference between
the major and intermediate axes, then $\xi_{|cr|}$ would vanish
and both $\xi_{|ar|}$ and
$\xi_{|br|}$ would equal $2/\pi\approx 0.64$).
The geometry of the environment is not strictly linear, especially
at larger scales.
However, despite these complications, the relationships
we have found provide useful quantitative predictions that can be tested when
full three-dimensional observations of halo shapes are available.

Previous studies have found that
the orientation of the major axis of a cluster is strongly
affected by the direction of the most recently accreted
subhalo, as seen in simulations \citep{vhvdw93,tormen97}
and observations \citep{ebd04}.
If these are accreted from filaments, then there should be
a strong alignment of cluster major axes with the
filaments, as we have found.
The minor axes are even more strongly aligned; this either
suggests that recently accreted subhalos have a stronger
effect on the minor axis than on the major axis,
or more likely that
figure rotation scrambles this alignment.
\citet{bs04-figrot}
found that most halos (78\%) show slow figure rotation
about the minor axis, with a median pattern speed of $\approx0.15~\hkmskpc$,
suggesting that the major axis of a typical halo can change by 90\degr\ in
a Hubble time. However, only 13\%\ of halos show figure rotation
about their major axis.
Therefore, even if a halo is formed with both its major and minor
axes perfectly aligned with the filament, the major axis most often
rotates away from that orientation by several degrees while the minor
axis most often remains in its original orientation.

The galaxy mass halos show weaker alignment between their
axes and the large scale structure than do the cluster mass halos.
It may be that the axes of lower mass halos are less affected
by recent major mergers, or that the direction of the accretion
is more isotropic \citep{vitvitska-etal02}.
\citet{apc04} find
that the accretion onto halos at masses down to
$5\times 10^{12}~M_{\Sun}$ is quite anisotropic; however,
97\%\ of our ``galaxy'' mass halos lie below this limit.
On the other hand, figure rotation may again be at work.
Although \citet{bs04-figrot} found no relationship between
the figure rotation pattern speed and the halo mass, lower mass
halos are on average dynamically older, and therefore figure rotation
may have had a longer time to modify the original orientation.
In any case, although we cannot directly test whether the
correlation of the axes depends on mass, the strong dependence
of axis alignment on mass suggests that the intrinsic alignments
of galaxy mass halos are much weaker than the intrinsic alignments
of cluster mass halos.
Therefore, when interpreting weak lensing results,
it is important to note that
the predictions for the intrinsic alignments of galaxies
made by analyzing cluster mass halos
\citep[e.g.][]{cm00,hrh00}
are likely overestimates.

Turning to the angular momentum, we find an intriguing correlation
between the concentration of a halo $c_{\mathrm{vir}}$ and
the tendency for its angular momentum to point toward other halos,
$\xi_{Lr}$
(see Figure~\ref{l dot r external all}). While most halos show
no tendency for the angular momentum to point toward rather than
away from nearby halos, halos with particularly high concentration
parameters show positive $\xi_{Lr}$.
The concentration
parameter is a measure of the factor by which the halo has
collapsed since its formation \citep{nfw97};
therefore non-linear effects and higher
order terms in the expansion of the potential are expected to play
a more important role in these halos.
Indeed, in simple
linear theory where the potential is 
approximated by a second-order Taylor expansion
\citep{white84}, each axis is symmetric and so it is impossible to
produce a preferred direction. However, a second-order expansion of
the potential $\Phi$ is only a solution to the Poission equation in
a constant density background. In order for the angular momentum to
point toward nearby structure, there must be a
background density gradient.
In the simplest case, the potential must be expanded to third order
to account for a linear density gradient.
Each additional
derivative of the potential couples to additional moments of the
density distribution
of the protohalo \citep{pdh02a}. Therefore, the third derivatives of
the potential couple to the skewness of the density distribution.
Because the third derivatives are no longer symmetric
(in general $T_{iij} \ne T_{ijj}$, where 
$T_{ijk} \equiv \partial^3 \Phi / \partial q_i \partial q_j \partial q_k$),
the symmetry is broken and it is possible to produce a preferred
direction.

The differences between the alignment of the angular momentum of low
and high mass halos are also intriguing.
Angular momentum is usually thought to either arise from the tidal
torquing of an asymmetric protohalo \citep{white84}, or by the accretion of
substructure on non-radial orbits \citep{tormen97,vitvitska-etal02}.
In a sense, these are not distinct scenarios;
accreted subhalos \textit{are} protogalactic material that has been tidally
torqued. However, the detailed orientation of the angular momentum
may depend on the clumpiness of the accretion and other
non-linear effects \citep{pdh02a}.

We find that the angular momentum in galaxy mass halos
points parallel to the large scale structure, while the angular momentum
in cluster mass halos points perpendicular
to the large scale structure (see Figure~\ref{l dot r abs}).
If the direction of the angular momentum is dominated by mergers
of objects falling in along filaments, then the angular momentum will
tend to point perpendicular to the velocity of the infalling
objects, i.e. perpendicular to the filament.
Therefore, the difference between the alignment
of the angular momentum of the low and high mass halos
suggests that the accretion onto low mass halos is either
smoother than the accretion onto high mass halos, or that a significant fraction
of the accretion in low mass halos occurs perpendicular to the filament
\citep{vitvitska-etal02}.
The group and cluster mass halos, on the other hand, tend to occur at the
intersections of large filaments, and have been built up by
the recent accretion of halos along the filament
which contribute angular momentum perpendicular to the filament
(note that this confirms the suggestion of \citet{mds02} that
accretion along a preferred axis is required to explain
the distribution of spin parameters seen in simulated halos).

This difference between the inferred origin of the angular momentum
of low and high mass halos
meshes nicely with that of \citet{peirani-etal04}.
These authors found that halos which have experienced a major merger acquired
their angular momentum much more rapidly than those halos which grew
only through steady accretion,
suggesting that the sources of their angular momenta are the merger events.
The halos in the merger sample also
built up their mass more quickly due to those mergers, and
consequently have larger median present-day masses.
Our group and cluster mass halos thus contain a larger fraction
of major merger products, and therefore a larger number of halos
whose angular momenta were built up through mergers.
It is encouraging that our results, based on the orientation of
the angular momentum vectors, and the results of \citet{peirani-etal04},
which are based on the evolution of the magnitude of the angular
momentum, build a consistent picture.

It is interesting to compare our results for collisionless dark
matter halos with the results of \citet{nas04} for simulations
that include baryonic physics. The main results of \citet{nas04} are
that the angular momenta of baryonic galactic disks
in gasdynamical simulations tend to align with
the intermediate axis of the local baryon distribution
on scales of $\approx 2~\hmpc$, and that observed edge-on
disk galaxies in the local supercluster have their spin axes
lying within the supergalactic plane, as expected in such a scenario.
We find that on the scale of an individual halo, the angular momentum
of the dark matter aligns with its minor axis. However,
we also find that it
lies parallel to the large scale structure, as expected if the spin
vector lies within local sheet-like structures as suggested by
\citet{nas04}.
We also find that in low density regions, halo spins point toward
nearby filaments, and are therefore aligned with the intermediate
axis of the local density field.
Therefore, despite the common presence of misalignments
between the angular momenta of the baryons and the dark matter in individual
galaxies \citep{ss04}, each retains some memory of the initial
torques provided by the large scale structure.

Unlike for galaxy mass halos, for which we
can confirm
that the angular momenta of the dark matter and baryons
share similar relationships to the large scale structure in simulations,
we do not have high resolution gasdynamical
simulations of the more massive groups and clusters with which to compare.
It may be possible to determine the rotation axis of baryons in
these systems observationally.
In theory, the spin vector of a relaxed cluster can be deduced from
the presence of a redshift gradient of the galaxies in the cluster;
however, confusion due to structure along the line of sight,
the small magnitude of the rotation compared to the intrinsic
velocity dispersion,
ambiguity in the orientation of the ellipsoidal shape,
intrinsic distance gradients,
the lack of a large sample of substructureless relaxed
clusters,
and the uncertain relationship between the angular momentum of the
cluster galaxies and that of the smooth X-ray emitting gas that dominates
the baryonic mass
make this measurement difficult.
The rotation of the X-ray gas itself, however, may be measured
using the kinematic Sunyaev-Zeldovich effect in future Cosmic
Microwave Background (CMB) surveys \citep{cm02,cc02}.

\section{Summary}\label{summary section}

We have studied the internal shapes
and angular momenta of galaxy and group mass
dark matter halos formed in a \lcdm\ $N$-body simulation,
and studied how they are correlated with the large scale
structure and the properties of neighbouring halos.

Internally, halos are triaxial with $b/a$ and $c/a$ ratios
of $0.75\pm0.15$ and $0.6\pm0.1$ respectively.
The distribution of axis ratios has a tail to low values.
The two-dimensional projected ellipticities cover
a broad range of values from 0 to 0.5, with a mean of 0.24,
consistent with the weak lensing results of \citet{hyg04}.
The axis ratios rise between 0.12 and 0.6~\rvir, beyond which
they drop. Within 0.12~\rvir, the measurement is probably
compromised by the force softening in the simulations.
Halos are most often prolate in the inner regions,
but tend to a more even mix of prolate and oblate at large radii.

The internal alignment of the halos within 0.6~\rvir\ is very
good, particularly for the minor axis, with a slight decrease in alignment
in the outermost regions.
High mass halos have particularly well-aligned axes.
While the orientation
of the angular momentum is also relatively constant,
it changes more noticeably as a function of radius than do the axes;
there is very little correlation between the very innermost regions and
the very outermost regions.
The distribution of spin parameters is essentially independent of the
size of the region used to calculate it, indicating that the
relative importance of the angular momentum is the same at
all radii.
At any given radius, the angular momentum vector tends to be
aligned with the minor axis and be perpendicular to the major axis.
However, the mean misalignment of $\sim 25\degr$ implies that galactic
disks are generally misaligned with their halo. Figure rotation
of the halo
and reorientation of the disk due to newly accreted angular momentum
act to maintain this misalignment.
The alignment between the angular momentum and the principal axes
gets weaker at larger radii.
The properties of the halo at 0.4~\rvir\ are quite characteristic
of their values at most other radii.

The minor axes of halos show a strong
tendency to lie perpendicular to the filaments.
As a consequence, the major (and, to a lesser degree, intermediate)
axes tend to point along filaments.
Figure rotation about the minor axis may be responsible for
the smaller degree of major axis alignment.
These alignments
fall off with distance as a power law.
In all cases, the alignment for group and cluster
mass halos is
stronger and extends to much larger separations than for galaxy mass halos.
Therefore, previous studies which predict the intrinsic alignments of galaxies
based on the shapes of cluster mass halos appear to systematically
overestimate the correlations.
The major axes
of halos show a weak correlation with those of other
nearby halos, while there is no robust detection of a correlation
for the intermediate or minor axes.

The angular momenta of high-concentration halos tend to point
toward, rather than away from, nearby halos. These halos have
collapsed further into the non-linear
regime, and therefore the effects of higher-order derivatives
of the tidal field, which are necessary to create a
preferential direction, are more important for these halos.
The angular momenta of halos in low density environments also tend to
point toward local density enhancements, in agreement with
the results of \citet{ko92} for galaxies 2~\hmpc\ or more
away from the local supergalactic plane.
The angular momenta of galaxy mass halos show a weak tendency to point along
filaments on scales up to 3~\hmpc, but those of group and cluster
mass halos show a very strong tendency to point perpendicular
to the filaments. This appears to be due to the different merger
histories of the two samples; higher mass halos are more likely
to have experienced a major merger
along a filament which dominated the evolution
of their angular momentum, while lower mass halos are more likely
to have had a smoother accretion history.
We detect no correlations between the angular
momentum directions of nearby halos, but due to the size of the
error bars, this is consistent with previous linear and $N$-body
studies that predict weak correlations.
Comparisons with recent gasdynamical simulations
and observations of edge-on disk galaxies in the local supercluster
suggest that
both the baryons and dark matter in galaxies share a memory of the
orientation of the large scale structure that provided the initial
torque.
The alignments we predict
may be tested with large samples of galaxy redshifts within
relaxed clusters, or by kinematic Sunyaev-Zeldovich studies in future
CMB experiments.

\acknowledgements
This work has been supported by grants from the U.S. National Aeronautics and
Space Administration (NAG 5-10827), the David and Lucile Packard Foundation,
the Bundesministerium f\"ur Bildung und Forschung (FKZ 05EA2BA1/8),
and by the American Astronomical Society and the
National Science Foundation in the form of an International Travel Grant.
 
\bibliography{ms.bib}
 
\end{document}